\documentclass[preprint,12pt]{elsarticle}

\usepackage{amssymb}
\usepackage{amsthm}
\usepackage{amsmath}

\usepackage{bm}
\usepackage{graphicx}
\usepackage{graphics}

\usepackage{subfigure}
\usepackage{tabularx,caption}
\usepackage{float} 
\usepackage{color}
\usepackage{multirow}
\usepackage{booktabs}
\usepackage{diagbox}
\usepackage{lineno}
\biboptions{sort&compress}

\usepackage[unicode]{hyperref}

\journal{Elsevier}

\begin{document}

\begin{frontmatter}



\title{Monte Carlo physics-informed neural networks for multiscale heat conduction via phonon Boltzmann transport equation}


\author[hust]{Qingyi Lin}
\author[hdu]{Chuang Zhang}
\author[hust]{Xuhui Meng \fnref{1}}
\author[hust]{Zhaoli Guo \fnref{1}}


\address[hust]{Institute of Interdisciplinary Research for Mathematics and Applied Science, School of Mathematics and Statistics, Huazhong University of Science and Technology, Wuhan 430074, China}
\address[hdu]{Department of Physics, School of Sciences, Hangzhou Dianzi University, Hangzhou 310018, China}

\fntext[1]{Corresponding authors: xuhui\_meng@hust.edu.cn (Xuhui Meng); zlguo@hust.edu.cn (Zhaoli Guo).}

\begin{abstract}

The phonon Boltzmann transport equation (BTE) is widely used for the description of multiscale heat conduction (from $nm$ to $\mu m$ or $mm$) in solid materials. Developing numerical approaches to solve this equation is challenging since it is a 7-dimensional integral-differential equation. 
In this work, we propose the Monte Carlo physics-informed neural networks (MC-PINNs), which do not suffer from the {\emph{``curse of dimensionality''}},  to solve the phonon Boltzmann transport equation for modeling the multiscale heat conduction in solid materials. In MC-PINNs, we utilize a deep neural network to approximate the solution to the BTE and encode the BTE as well as the corresponding boundary/initial conditions using the automatic differentiation. In addition, we propose a novel two-step sampling approach  to address the issues of inefficiency and inaccuracy in the widely used sampling methods in PINNs. In particular, we first randomly sample a certain number of points in the temporal-spatial space (Step I) and then draw another number of points randomly in the solid angular space (Step II). The training points at each step are constructed based on the data drawn from the above two steps using the tensor product. The two-step sampling strategy enables the MC-PINNs (1) to model the heat conduction from ballistic to diffusive regimes, and (2) to be more memory efficient compared to the conventional numerical solvers or existing PINNs for BTE. A series of numerical examples including quasi-one-dimensional (quasi-1D) steady/unsteady heat conduction in a film, and the heat conduction in quasi-two- and three-dimensional square domains, are conducted to justify the effectiveness of the MC-PINNs for heat conduction spanning diffusive and ballistic regimes.  Finally, we perform a comparison on the computational time and the memory usage between the MC-PINNs and one of the state-of-the-art numerical methods to demonstrate the potential of MC-PINNs for large-scale problems in real-world applications.

\end{abstract}



\end{frontmatter}


\section{Introduction}
\label{sec:sec1}

Microdevices, e.g., microprocessors and microelectronic circuits, often generate heat with great density during operation. Efficient thermal management is crucial to the reliability and optimal performance of these devices. As reported in \cite{stettler2021industrial}, the heat conduction in microdevices like chips is inherently a multiscale problem spanning nanoscale (1-100 nm) to microscale (0.1-100 $\mu$m) or macroscopic (0.1-100 mm). The classical Fourier's law, which is a widely used empirical {\emph{macroscopic}} model for heat conduction in solids,   is not capable of describing the heat conduction in the non-diffusive or ballistic regime  \cite{chen2021non,zhang2007nano, cahill2014nanoscale}. Nevertheless, the phonon Boltzmann transport equation (BTE) is a well-established model for the descriptions of the heat conduction in solid materials spanning the ballistic and diffusive regimes \cite{chang2023calibrated,stettler2021industrial,ziman2001electrons,murthy2005review}. 


To advance the understanding of multiscale heat conduction in solid materials, lots of efforts have been taken into developing numerical methods for solving the phonon BTE. However, developing effective numerical methods to efficiently solve the phonon BTE is challenging since it is a 7-dimensional integral differential equation in real-world applications. The conventional numerical methods can be broadly categorized into two main approaches, i.e., the stochastic methods, e.g., direct simulation Monte Carlo (DSMC)~\cite{peraud2011efficient,mazumder2001monte,jeng2008modeling,lacroix2005monte}, etc., and deterministic approaches, e.g., discrete velocity/ordinate method~\cite{sellan2010cross,nabovati2011lattice, guo2016lattice, narumanchi2005comparison}, synthetic iterative scheme~\cite{zhang2017unified,zhang2019implicit,zhang2023acceleration}, discrete unified gas kinetic scheme (DUGKS)~\cite{guo2013discrete,guo2016discrete}, and so on. The former uses discrete meshes in the temporal-spatial space and is efficient for steady problems in the ballistic regime. However, it converges slowly for problems in the (near) diffusive regime as well as non-stationary problems.  The latter employs the discrete meshes in the temporal-spatial-angular space and is able to achieve good accuracy for steady/unsteady problems spanning a wide range of transport regimes. In general, the number of discrete points in the solid angular space required to obtain accurate results for the heat conduction problem in the diffusive regime is much smaller than in the ballistic regime. The requirement of large numbers of discrete points in the solid angular space to avoid the {\emph{ray effect}} in the ballistic regime ~\cite{lathrop1968ray,chai1993ray} leads to intensive computational cost and prohibitive memory usage for large-scale thermal engineering problems in practice.


Deep neural networks (DNNs), which are able to break the {\emph{curse of dimensionality}}, have recently emerged as an effective tool for solving partial differential equations (PDEs), such as physics-informed neural networks (PINNs)~\cite{raissi2019physics,raissi2020hidden,lu2021physics, cuomo2022scientific}, DeepRitz~\cite{yu2018deep}, and mothods combining moment closure with neural networks~\cite{ly2007critical,huang2022machine,krupansky2023deep}. Generally, the DNN-based PDE solvers are mesh-free, which saves significant effort for generating the meshes in conventional methods, e.g., finite volume method, and finite element method. Among the aforementioned deep learning approaches, the PINNs, which encode the PDEs via the automatic differentiation technique, are one of the most widely used methods for solving PDE problems due to their effectiveness and straightforward implementations. 
Recently, the PINNs have been successfully employed to simulate flows ranging from continuous to rarefied flows by solving the Boltzmann-BGK or Boltzmann equation~\cite{de2022physics, oh2024separable}, demonstrating their capability of handling multiscale problems~\cite{lou2021physics,jin2023asymptotic}. Inspired by the aforementioned work for modeling multiscale flows using PINNs, Li {\sl et al.} proposed to solve the phonon BTE using PINNs to study the heat conduction in solid materials~\cite{li2021physics, li2022physics}. The results showed that the PINNs are capable of modeling heat conduction ranging from ballistic to diffusive regimes with satisfactory accuracy. In particular, the solid angular space are discretized as a priori for training the PINNs in previous works~\cite{lou2021physics,li2021physics,li2024solving}.  For instance, the angular space is discretized utilizing the points in the Gauss Legendre quadrature rule in \cite{li2021physics,li2024solving}, which is similar to the way that is widely used in the conventional deterministic approaches for solving phonon BTE, e.g., DUGKS. 
Although the effort of generating the meshes in temporal-spatial space can be saved in the PINN approaches, it shares the same defect as in the conventional deterministic approaches, i.e., a large number of discrete points in the angular space is required for heat conduction at the ballistic regime. Moreover, for scenarios in real-world applications where the heat conduction at the micro- and nanoscale coexists,  a large number of discrete points in the angular space of BTE is also required for the accurate descriptions of the non-Fourier's behavior at the ballistic regime, since it is challenging to develop adaptive discretization for heat conduction at different scales. Therefore, we still have difficulty employing the PINN methoeds developed in \cite{li2021physics,li2024solving} for large-scale heat conduction problems due to the expensive computational as well as memory cost. Numerical approaches for solving the phonon BTE that are more flexible as well as efficient are herein still desirable.

In this study, we employ the PINNs as the backbone to solve the phonon BTE due to their effectiveness and capability for handling high-dimensional problems. In particular, we propose a novel two-step sampling approach for the training of PINNs to address the issue of expensive computational and memory costs in the existing PINN models. We refer to the PINNs equipped with the proposed two-step sampling approach as the Monte Carlo PINNs (MC-PINNs), which (1) is capable of modeling the heat conduction spanning ballistic to diffusive regimes, (2) is mesh-free in the temporal-spatial as well as the angular space, (3) is a unified approach at all regimes without the requirement to employ different discretizations in the angular space for heat conduction at different scales, and (4) requires much less discrete points in the angular space at each training step, and thus is more computationally and memory efficient comparing to the approach in \cite{li2021physics,li2024solving}.


The rest of the paper is organized as follows. In Sec.~\ref{sec:method}, we introduce the methodology, including the phonon BTE, and the PINN approach along with the two-step sampling strategy which is crucial for multiscale heat conduction. In Sec.~\ref{sec:results}, a series of numerical experiments are conducted, including steady and unsteady quasi-1D, quasi-2D, and 3D multiscale heat conduction problems. We finally summarize this study in Sec.~\ref{sec:summary}.

\section{Methodology}\label{sec:method}

In this section, we first introduce the phonon Boltzmann transport equation as well as the corresponding boundary/initial conditions, which are used to describe the heat conduction from the ballistic to diffusive regime. We then present the MC-PINNs, i.e., physics-informed neural networks, and the proposed two-step sampling approach for the training of PINNs.  

\subsection{Phonon Boltzmann transport equation}


The phonon BTE is widely used for simulating heat conduction in solid materials at different scales. Various models of the phonon BTE have been developed to address different applications~\cite{mazumder2001monte}. For instance, in~\cite{chiloyan2021green}, the full/linearized phonon BTE is utilized to consider the complex interactions among different phonon modes;  
the frequency-resolved BTE  is often employed in applications that require to consider waves with specific frequencies~\cite{jeng2008modeling,lacroix2005monte,zhang2019implicit,zhang2023acceleration}.
Further, the gray model is generally used for multiscale heat conduction in three dimensions subject to minor temperature variations~\cite{guo2016discrete,zhang2017unified,zhang2023acceleration}.



\begin{figure}
    \centering
    \includegraphics[width=0.8\textwidth]{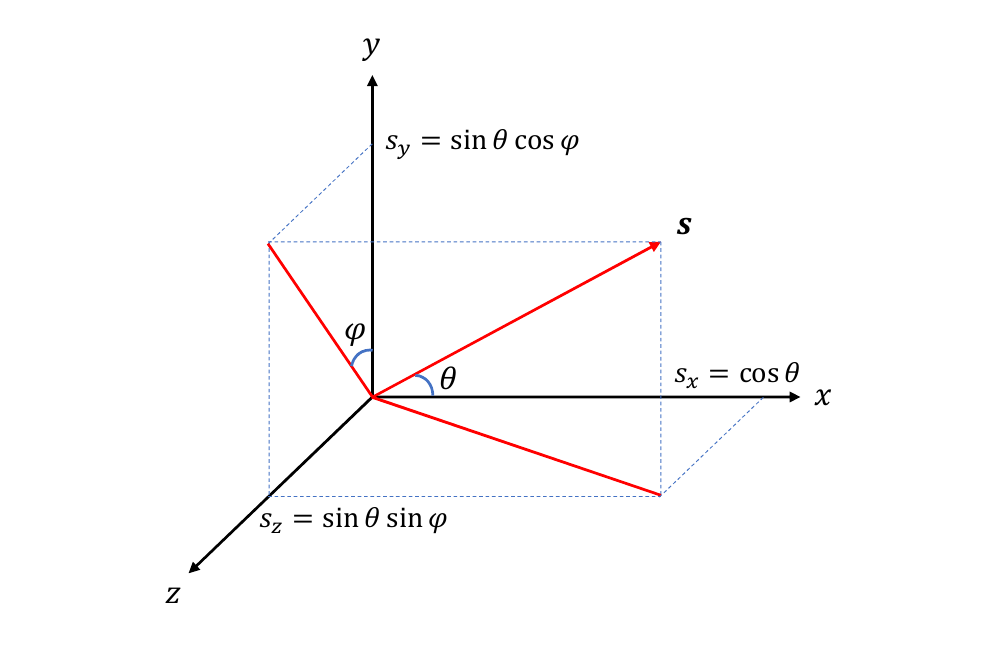}
    \caption{Schematic of the unit directional vector in solid angular space with polar angle denoting by the polar angle $\theta$ and azimuthal angle $\varphi$. $s_x$, $s_y$ and $s_z$ are the projections of the directional
    vector in Cartesian coordinates.}
    \label{fig:coordinate}
\end{figure}

In this study, our particular interest is on the phonon gray model, which is specifically developed for probing multiscale heat conduction in three-dimensional materials with minor temperature variations. 
Given a certain number of reference variables, e.g.,  temperature $T_{ref}$,  heat $C_{ref}$,  length $L_{ref}$, group velocity $v_g$ and time $t_{ref}=L_{ref}/v_g$,  the dimensionless phonon BTE in the gray model can then be obtained as
\begin{subequations}\label{eq:bte}
\begin{equation}\label{eq:bte_a}
\frac{\partial e^{\prime\prime} }{\partial t}+  \boldsymbol{s}\cdot\nabla_{\bm{x}} e^{\prime\prime} =\frac{1}{\text{Kn}}(e^{eq}-e^{\prime\prime}),  
\end{equation}
\begin{equation}\label{eq:bte_b}
e^{eq} = \frac{C_V  T }{4 \pi},
\end{equation}
\end{subequations}
where $e^{\prime\prime}$ represents the phonon distribution function of energy density, which is a function of time $t$, spatial coordinate $\bm{x}$, and the unit directional vector $\bm{s}$.
The directional vector is uniformly distributed on the surface of a unit sphere in the phonon BTE, and thus can be expressed in the Cartesian coordinates as:
\begin{equation}
    \bm{s}=(s_x, s_y, s_z)=(\cos\theta, \sin\theta \cos\varphi, \sin\theta \sin\varphi),
\end{equation}
where $\theta\in[0, \pi]$ and $\varphi\in[0, 2\pi]$ are the polar and azimuthal angles, respectively. 
The schematic of the directional vector in solid angular space with polar and azimuthal angle is shown in Fig.~\ref{fig:coordinate}.
In addition, $e^{eq}$ denotes the equilibrium distribution function for the energy density; $C_V$ and $T$ are the specific heat and temperature, respectively.  Moreover, $\text{Kn} = \lambda/L_{ref}$ is the Knudsen number defined as the ratio between the phonon mean free path to the characteristic length of a specific system.
Due to the law of energy conservation, we can obtain the  following equation for the right-hand side of Eq. \eqref{eq:bte_a}: 
\begin{align}\label{eq:t_q}
0=  \int_{4 \pi} \frac{e^{eq}-e^{\prime\prime}}{\text{Kn}} d\Omega,
\end{align}
where $d\Omega$ represents the integral over the {surface of the unit sphere}.
The quantities of interest (QoI) in heat conduction problems including temperature $T$ and heat flux $\bm{q}$ can be obtained by taking the moment of phonon energy density distribution function, i.e.,
\begin{align}
T = \frac{\int e^{\prime\prime} d\Omega}{C_V}, ~~~ \bm{q} = \int  \bm{s} e^{\prime\prime} d\Omega.
\end{align}


Different boundary conditions are generally required to solve Eq. \eqref{eq:bte} in different applications. For example, the diffuse reflection conditions can be used for multiscale heat conduction with rough surfaces \cite{zhang2017unified}; and the periodic boundary condition is often employed in scenarios with periodic geometries.  In the current study, we adopt the thermalization boundary condition, in which we assume that the temperatures at the boundaries are known~\cite{fryer2014moment}. For this particular case, the phonons are absorbed upon striking the boundary, and new phonons with equilibrium states are emitted. 
We note that the aformentioned boundary condition is equivalent to imposing the  Dirichlet boundary condition for the phonon energy density function $e^{\prime\prime}$, which is expressed as follows:
\begin{equation}\label{eq:bc}
e^{\prime\prime}(t,x_w, \boldsymbol{s}) =\frac{C_V T_w }{4\pi},\quad \boldsymbol{s}\cdot \boldsymbol{n}>0,
\end{equation}
where $x_w$ refers to the boundary, $\boldsymbol{n}$ is the unit normal vector pointing from the boundary to the computational domain, and $T_w$ denotes the temperature at the boundary.
For initial conditions, we assume that the distribution function is at the equilibrium state with the temperature at the initial time, which reads as \cite{guo2016discrete}:
\begin{equation}\label{eq:bc}
e^{\prime\prime}(t_0,\boldsymbol{x}, \boldsymbol{s}) =\frac{C_V T_0 }{4\pi},
\end{equation}
where $T_0$ denotes the temperature at the initial time $t_0$.

\begin{figure}[H]
    \centering
    \includegraphics[width=0.9\textwidth]
    {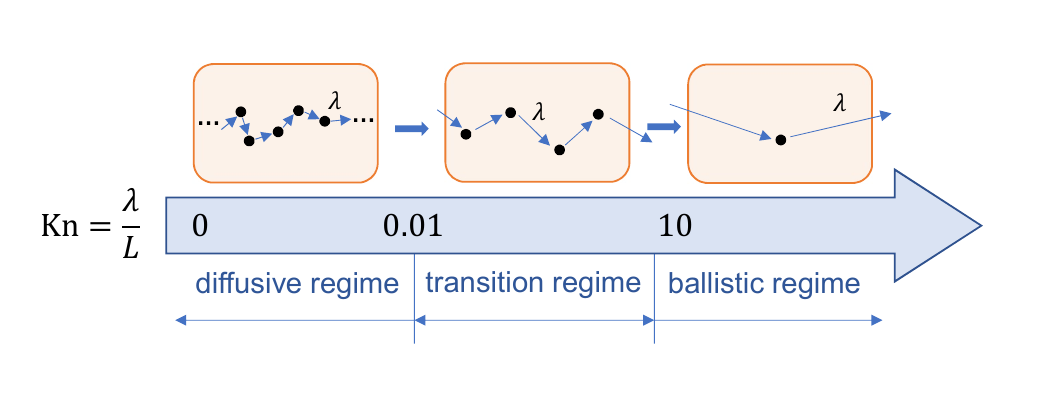}
    \caption{Schematic of transport regimes based on the Knudsen number. $L$ is the characteristic length of the system and $\lambda$ is the phonon mean free path.}
    \label{fig:regime}
\end{figure}

We would like to discuss that the multiscale heat conduction is generally divided into different transport regimes based on the Knudsen number, as shown in Fig.~\ref{fig:regime}.
For $\rm{Kn}<0.01$, the characteristic length of the system is much larger than the phonon mean free path, and the phonon transport is in the diffusive regime. The temperature field is generally described by the  Fourier's law. As $\rm{Kn}$ increases, the Fourier's law begins to fail. For $\rm{Kn}>10$, the heat conduction is dominated by the ballistic effects.

\subsection{PINNs for phonon BTE}

\subsubsection{Physics-informed neural networks}
\label{sec:pinn_bte}

\begin{figure}[H]
    \centering
    \includegraphics[width=1.0\textwidth]
    {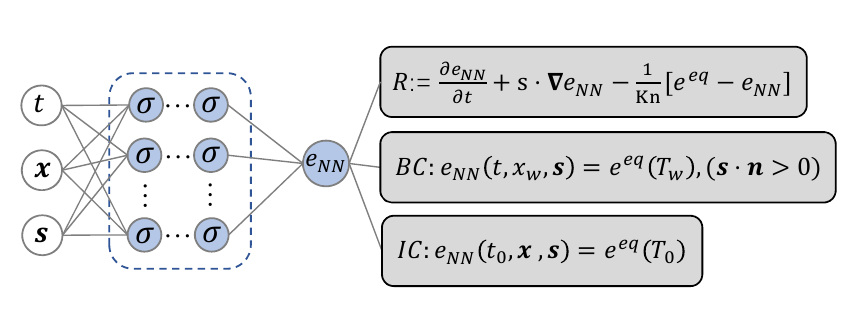}
    \caption{Schematic of physics-informed neural networks for phonon Boltzmann transport equation. $(t, \boldsymbol{x}, \boldsymbol{s})$ is the input, $\sigma$ are the activation functions in the networks, and $e_{NN}$ is the output of the neural networks. The gray boxes represent the physics-informed parts, i.e., the governing equation and boundary/initial conditions.}
    \label{fig:PINNs}
\end{figure}

The PINNs for solving phonon BTE are illustrated in Fig. \ref{fig:PINNs}. As shown, we first utilize DNNs, which take inputs as $(t, \bm{x}, \bm{s})$, to approximate the solution to Eq. \eqref{eq:bte}, i.e., $e^{\prime \prime}$, and then the governing equation, as well as the boundary and/or the initial conditions, are encoded in the DNNs via the automatic differentiation \cite{baydin2018automatic}. The loss function for training the PINNs is composed of the mean squared errors for the residual of the equation, mismatches between the predictions of DNNs and the boundary/initial conditions, which is expressed as follows:
\begin{equation}
	\mathcal{L} = \mathcal{L}_{Eq} + \mathcal{L}_{BC} + \mathcal{L}_{IC},
\end{equation}
where $\mathcal{L}$ is the total loss, $\mathcal{L}_{Eq}$, $\mathcal{L}_{BC}$, and $\mathcal{L}_{IC}$ denote the losses corresponding to the loss of equation, boundary, and initial conditions, respectively. Specifically, $\mathcal{L}_{Eq}$, $\mathcal{L}_{BC}$ and $\mathcal{L}_{IC}$ are computed following
\begin{equation}\label{eq:loss}
	\begin{aligned}
	   \mathcal{L}_{Eq}&=\frac{1}{N}\sum^{N}_{i=1}|R_i(t, \bm{x}, \bm{s})|^2, \\
		\mathcal{L}_{BC}&=\frac{1}{N_{BC}}\sum^{N_{BC}}_{j=1}|e_{NN,j}(t, \bm{x}_w, \bm{s})-e^{eq}_j(t, \bm{x}_w, \bm{s})|^2,\\
      \mathcal{L}_{IC}&=\frac{1}{N_{IC}}\sum^{N_{IC}}_{k=1}|e_{NN,k}(t_0, \bm{x}, \bm{s}) - e_k^{eq}(t_0, \bm{x}, \bm{s})|^2,\\
	\end{aligned}
\end{equation}
and
\begin{equation}
R={\partial_t e}_{NN}+\boldsymbol{s}\cdot\nabla e_{NN} - \frac{1}{\rm{Kn}}(e^{eq}- e_{NN}),
\end{equation}
where $R$ denotes the residual of the equation,
$N$, $N_{BC}$ and $N_{IC}$ are the numbers of training points used to compute the loss functions in PINNs for the residual, boundary, and initial conditions, respectively; $R_i$, $e^{eq}_j/e_{NN,j}$ and $e^{eq}_k/e_{NN,k}$ are quantities evaluated at points $(t, \bm{x}, \bm{s})_i$ for $i = 1, ..., N$; $(t, \bm{x}_w, \bm{s})_j$ for $j = 1, ..., N_{BC}$; and $(t_0, \bm{x}, \bm{s})_k$ for $k = 1, ..., N_{IC}$, respectively.
In the present study, we refer to the points used to evaluate $\mathcal{L}_{Eq}$, $\mathcal{L}_{BC}$, and $\mathcal{L}_{IC}$ as residual, boundary, and initial points, respectively.

We note that the computations of the residual for the equation, i.e., $R_i$, require the calculation of the equilibrium energy density function, which is an integral defined in Eq. \eqref{eq:t_q}. Different approaches can be utilized to compute the integral since we can obtain the prediction of the energy density function at an arbitrary point $(t, \bm{x}, \bm{s})$ in the temporal-spatial-angular space via the DNNs, e.g., Monte Carlo method, Gauss-Legendre, and Newton-Cotes quadrature. In the present study, it is found that Monte Carlo method, which is computationally efficient at both low and high dimensions, is able to provide results with satisfactory accuracy. Hence, the Monte Carlo method is employed for the computations of $e^{eq}$ in this work.

As for the selection of the training points for evaluating the loss function in PINNs (Eq. \eqref{eq:loss}), two different ways are often used, i.e., (1) the number ($N/N_{BC}/N_{IC}$) as well as the locations of the training points are prescribed at each training step, and (2) the number of points ($N/N_{BC}/N_{IC}$) is prescribed, but the corresponding locations are randomly generated at each training step.  The former is similar to the meshes in the conventional numerical methods. As mentioned, lots of discrete points in the solid angular space are required to accurately capture the non-Fourier behavior at the ballistic regime, which results in expensive computational and memory costs for multiscale heat conduction problems in real-world applications.  
For the latter approach, if we randomly generate training points in the temporal-spatial-angular domain, denoted as $(t, \bm{x}, \bm{s})_i, i = 1, ..., M$, it is probable that each point $(t, \bm{x}, \bm{s})_i$ will be different for different $i$. As a result,  we only solve the energy density function corresponds to $\bm{s}_i$ at $(t, \bm{x})_i$ in each training step. It leads to slow convergence or even degenerated results since we need to compute the integral in the entire angular domain, as we will show in Sec. \ref{sec:1d}.

\subsubsection{Two-step sampling approach}
\label{sec:two_step}

To address the issues mentioned in the above two sampling methods, we propose the following two-step sampling strategy at each training step, i.e., (1) we first randomly draw $B_{t, \bm{x}}$ points in the temporal-spatial domain, i.e., $(t, \bm{x})_i, ~i = 1, ..., B_{t, \bm{x}}$ (Step I); and (2) we then randomly draw $B_{\bm{s}}$ points in the solid angular domain, i.e., $\bm{s}_j,~ j = 1, ..., B_{\bm{s}}$ (Step II). The training points in the temporal-spatial-angular space $(t, \bm{x}, \bm{s})$ at each training step are constructed using the tensor product given the points sampled at the two steps above.  In general, we set $B_{\bm{s}} > 1$ to improve the accuracy for the computations of the integral in Eq. \eqref{eq:t_q} as well as $e^{eq}$.  Hence, we expect that the proposed two-step sampling approach is able to achieve better accuracy when compared to the second sampling method mentioned in Sec. \ref{sec:pinn_bte}.  Note that for steady problems, we can just drop the dimension $t$, and perform the two-step sampling approach only in the spatial and angular domains.  In our computations, to obtain the samples of $\bm{s}$,  we first randomly sample the projection of the polar angle $\cos\theta \in [-1, 1]$ along the $x$-axis, and sample the azimuthal angle $\varphi \in [0, 2\pi]$ to generate the random points in the solid angular space.

For the implementations of boundary/initial conditions, we  randomly generate a certain number of training points in the temporal-spatial-angular domain, i.e., $B_{t,\bm{x},\bm{s}}$, since we do not need to compute the integral in boundary/initial conditions, for computational efficiency.  


Here we further discuss the superiority of the two-step sampling approach over the the first sampling method in Sec. \ref{sec:pinn_bte}. As aforementioned, a large number of points in the solid angular domain have to be employed in the ballistic regime 
for the first approach of Sec. \ref{sec:pinn_bte}. In addition, how to determine the optimal number of discrete points at different regimes remains an open question. In the two-step sampling method, we randomly generate a certain number of points in the solid angular space at each training step. We can then traverse the entire angular domain even if we employ a small number of points in each step as long as we have enough training steps, which is similar to the statistic or particle methods for solving the BTE. Hence, compared to the discrete ordinate/velocity methods, the present solver is {\emph {free of ray effects}} in the ballistic regime. Therefore, less computational and memory cost for heat conduction at the ballistic regimes are then expected at each training step if the two-step sampling strategy is adopted. As we will show in Sec. \ref{sec:1d}, the two-step sampling method is more flexible than the first sampling approach in Sec. \ref{sec:pinn_bte} for multiscale heat conduction problems.   Theoretical analyses for providing intuition on the convergence of the two-step sampling approach is further present in~\ref{sec:appendix_a} following~\cite{garrigos2023handbook}. 



\section{Results and Discussion}\label{sec:results}

In this section, we test the proposed method for simulating the heat conduction in solids ranging from diffusive to ballistic regimes, using four demonstration examples: (1) steady and unsteady quasi-one-dimensional (quasi-1D) film heat conduction, (2) quasi-two-dimensional (quasi-2D) heat conduction in a square domain, and (3) three-dimensional (3D) cuboid phonon transport.  Further, the Adam optimizer, which is a variant of the stochastic gradient descent method, is utilized to minimize the loss functions in MC-PINNs. The details for computations, e.g., the architectures of NNs,  activation functions, and the numbers of points used in Monte Carlo method/training steps, etc., as well as the details for reference solutions, are present in \ref{sec:appendix_b}. 


\subsection{Quasi-1D film heat conduction}

\begin{figure}[H]
    \centering
    \includegraphics[width=0.5\textwidth]{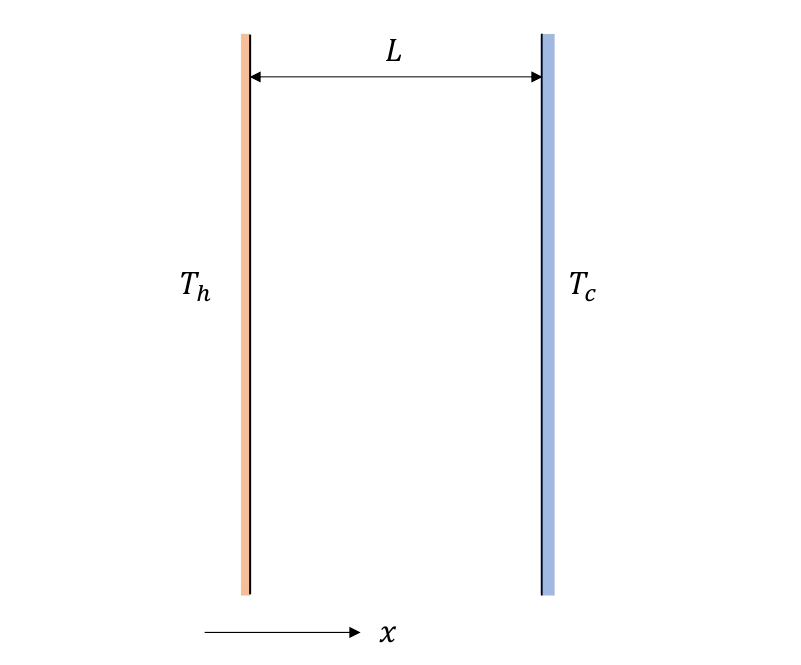}
    \caption{Schematic of the quasi-1D film heat conduction problem. $T_h$ and $T_c$ are the temperatures at the left and right boundaries, respectively;  and $L$ is the thickness of the film.}
    \label{fig:film}
\end{figure}

We first consider a quasi-1D heat conduction problem across a film, as depicted in Fig.~\ref{fig:film}. The thickness of the film is $L=1$, which serves as the characteristic length here. The temperatures at the left and right boundaries, i.e., $x = 0 $ and $L$, are $T_h = 1$ and $T_c = 0$, respectively. We employ the MC-PINNs to model the heat conduction at four different Knudsen numbers, i.e., $\rm{Kn} = 0.01, 0.1, 1$, and $10$, to justify the capability of the present method for multiscale heat conduction.  For the quasi-1D problem, the angular variable can be expressed as $s_x=\cos{\theta}$. In all test cases, we randomly select $B_{\bm{x}} = 40$ and $B_{\bm{s}} = 16$ points in the spatial and angular domain for the computations of the loss of the equation at each training step, respectively. For the boundary conditions, we randomly select $B_{x_w, \bm{s}} = 50$ points in the solid angular space at the boundary $x = 0$ and $x=1$ to implement the thermalization boundary condition in Eq.~\eqref{eq:bc}.

\label{sec:1d}

\begin{figure}[H]
\centering
\subfigure[]{\includegraphics[width=0.48\textwidth]{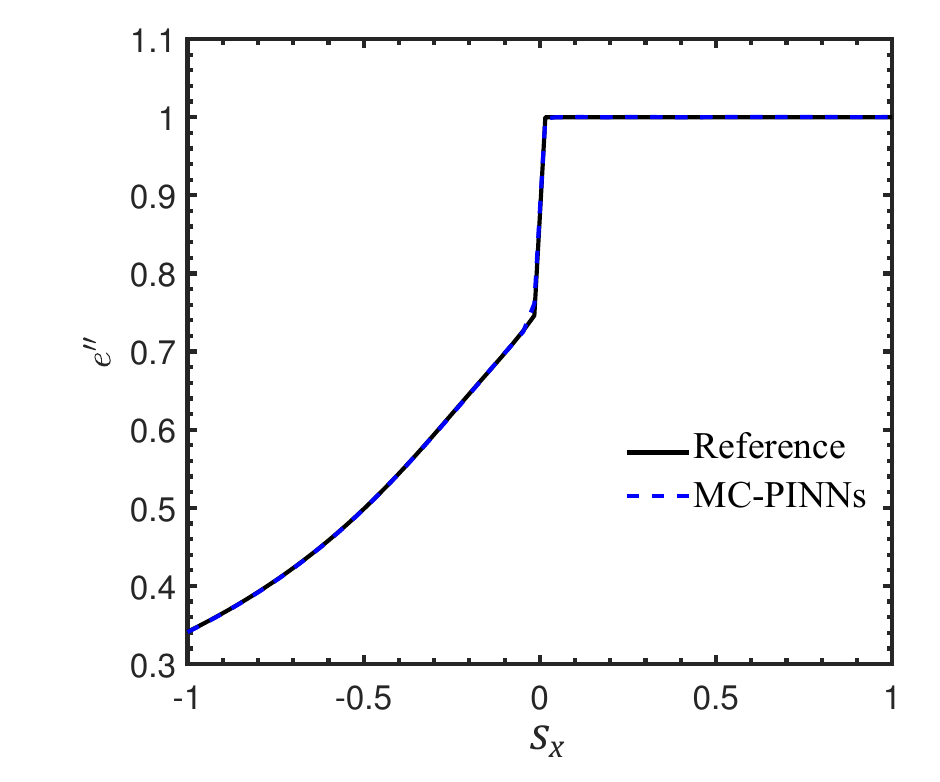}\label{FIG:f1}}
\subfigure[]{\includegraphics[width=0.48\textwidth]{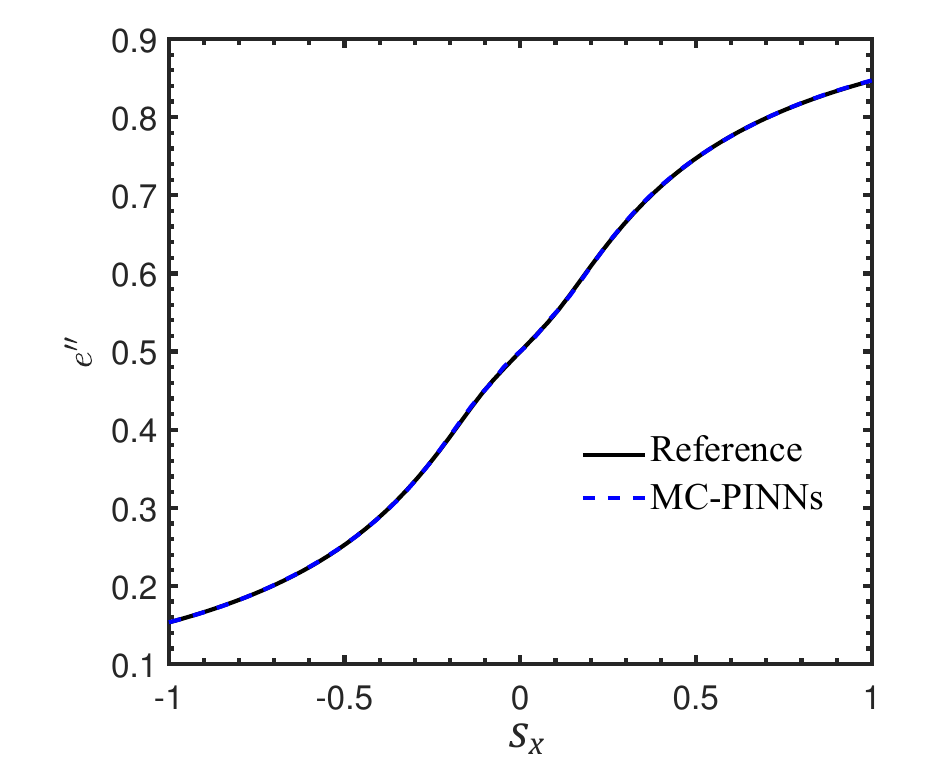}\label{FIG:f2}}
\caption{
Quasi-1D film heat conduction: Predicted energy distribution function from MC-PINNs at $\rm{Kn}=1$ for (a) $x = 0$, and (b) $x = 0.5$.  Black solid line: Reference solution \cite{guo2016discrete}; Blue dashed line: MC-PINNs.
%
\label{FIG:f}
}
\end{figure}

\begin{figure}[h]
\centering
\subfigure[]{\includegraphics[width=0.48\textwidth]{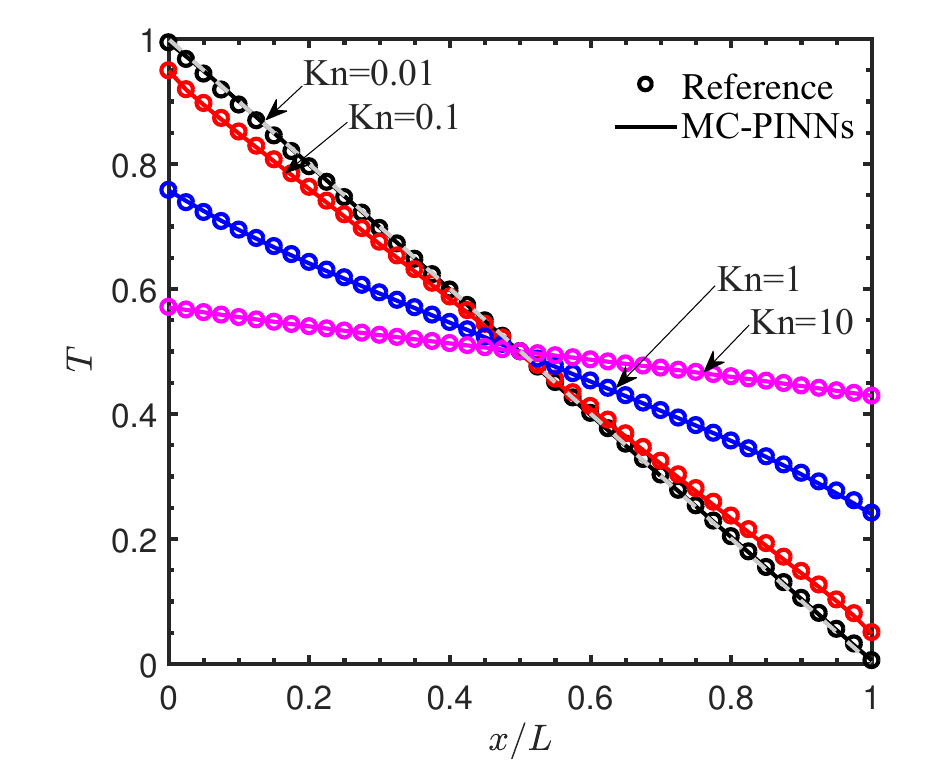}\label{fig:FIG_1d_T}}
\subfigure[]{\includegraphics[width=0.48\textwidth]{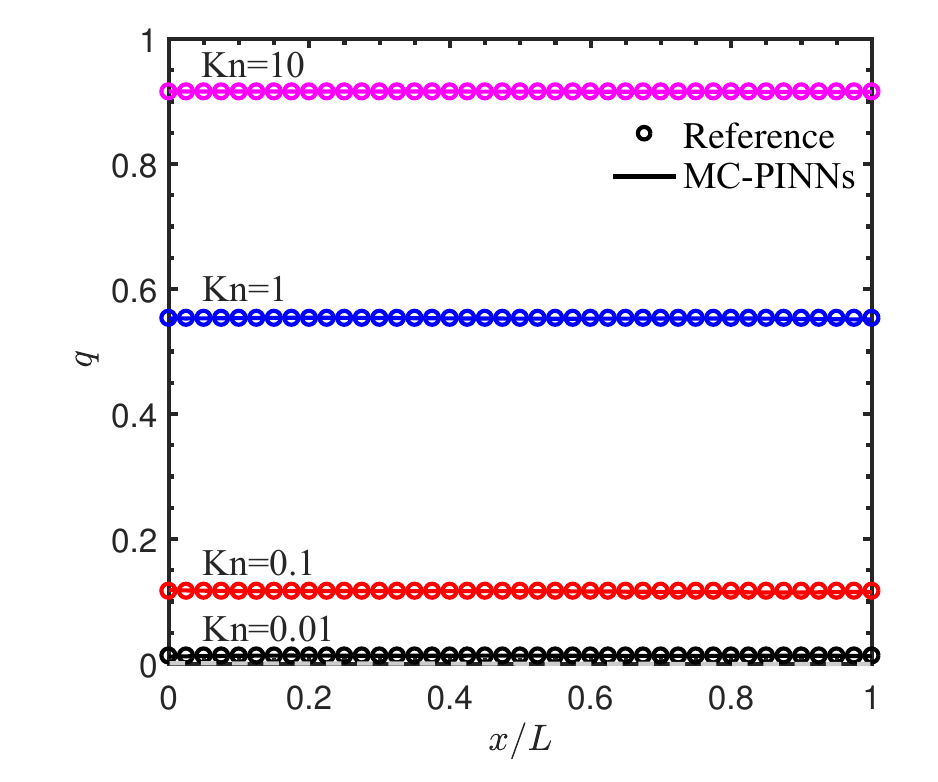}\label{fig:FIG_1d_q}}
\caption{
Quasi-1D film heat conduction: Predicted (a) temperature,  and (b) heat flux, from MC-PINNs at different $\rm{Kn}$. Symbols: Reference solution \cite{zhang2019implicit}; Solid line: MC-PINNs;  Gray dashed line: the solution of Fourier's Law.
\label{fig:FIG_1d_1}}
\end{figure}

To justify the accuracy of the present method, we first present the energy distribution functions computed from the MC-PINNs for the case with $\rm{Kn} = 1$ at two representative locations, i.e., $x = 0$ and $0.5$,  in Fig. \ref{FIG:f}. As shown, the results from the MC-PINNs agree well with numerical results in~\cite{guo2016discrete}.  In addition, the temperature as well as the heat flux are QoI in heat conduction problems rather than the energy distribution functions. We then illustrate the predicted temperature and the heat flux for $x \in [0, 1]$ from MC-PINNs in Fig. \ref{fig:FIG_1d_1}. It is observed that: (1) all the results from the MC-PINNs are in good agreement with those in \cite{zhang2019implicit}, (2) the temperature exhibits significant nonlinearities near the boundaries for the test cases with $\rm{Kn} > 0.01$ (Fig. \ref{fig:FIG_1d_T}), and (3) the heat flux $q$ remains constant although the temperature is nonlinear for $\rm{Kn} > 0.01$, which is clearly different from the widely used Fourier's law (Fig. \ref{fig:FIG_1d_q}). In other words, the heat conduction for cases with $\rm{Kn} > 0.01$ exhibits significant non-Fourier behavior.  

To further demonstrate the accuracy of the present method, we display the relative $L_2$ errors between the predicted temperature and heat flux from MC-PINNs and the reference solutions\cite{guo2016discrete} in Table~\ref{tab1}, where the relative $L_2$ error is computed as follows: 
\begin{equation}\label{eq:l2}
E_{\phi} = \frac{\Vert \phi_{ref} - \phi_{PINNs} \Vert_2}{\Vert \phi_{ref}\Vert_2},~ \phi = T, q.
\end{equation}
As shown, the relative $L_2$ errors between the present method and the reference solutions for all test cases are less than $1\%$, demonstrating that the MC-PINNs are capable of simulating heat conduction at both diffusive and ballistic scales with good accuracy.  Additional results on the effect of $B_{\bm{s}}$, i.e., the number of points used at each training step in the angular domain, on the predicted accuracy are present in \ref{sec:appendix_c}.

\begin{table}[H] 
\caption{Quasi-1D film heat conduction: Relative $L_2$ errors between MC-PINNs and reference solution at different $\rm{Kn}$.
$E_{\phi}$: Relative $L_2$ errors for temperature and heat flux.
\label{tab1}}
\newcolumntype{C}{>{\centering\arraybackslash}X}
\begin{tabularx}
{\textwidth}{CCCCC}
\toprule
 - & $\rm{Kn}=0.01$	& $\rm{Kn}=0.1$ & $\rm{Kn}=1$ & $\rm{Kn}=10$\\
\midrule
$E_T$			& 0.515\% & 0.107\%& 0.052\% & 0.142\%\\
$E_q$		& 1.000\% &0.162\% & 0.090\%& 0.743\% \\
\bottomrule
\end{tabularx}
\end{table}

As mentioned, the selection of training points in the solid angular space is crucial for the simulations of multiscale heat conduction. We now study the effect of different sampling approaches in the angular space on the predicted accuracy. Specifically, we test the following five different sampling approaches to obtain the residual points for training the PINNs:
\begin{enumerate}
    \item \textbf{PINNs-1:} The two-step sampling proposed in the present study: we randomly sample $B_{\bm{x}}$ points in the spatial domain and $B_{\bm{s}}$ points in the solid angular domain, and then use a tensor product to generate the residual points at each training step.
    \item \textbf{PINNs-2:} We employ $B_{\bm{x}}$ and $B_{\bm{s}}$ uniform discrete points in the spatial and solid angular domains, respectively. The tensor product is then utilized to generate the residual points at each training step, simialar as in PINNs-1. 
    \item \textbf{PINNs-3:} In the spatial domain, we use $B_{\bm{x}}$ uniform discrete points, and $B_{\bm{s}}$ quadrature points from the Gauss-Legendre method are employed in the solid angular domain, The residual points at each training step is then constructed using the tensor product. We note that this sampling approach is the same as in \cite{li2021physics}.
    \item \textbf{PINNs-4:} We randomly sample $B_{\bm{s}}$ points in the solid angular domain, but use $B_{\bm{x}}$ uniform points in the spatial domain. The tensor product is then use to generate the residual points at each training step, simialar as in PINNs-1. Note that the sampling approach here can be viewed as a special case in PINNs-1 by replacing the random points with the fixed uniform grids in the spatial domain. 
    \item \textbf{PINNs-5:} Randomly sample $B_{\bm{x}} \times B_{\bm{s}}$ residual points in the spatial-angular domain at each training step. We note that this approach is one of the most widely used sampling methods in PINNs. 
\end{enumerate}

\begin{figure}[h]
\centering
\subfigure[]{\label{fig:1d_methods_a}
\includegraphics[width=0.3\textwidth]{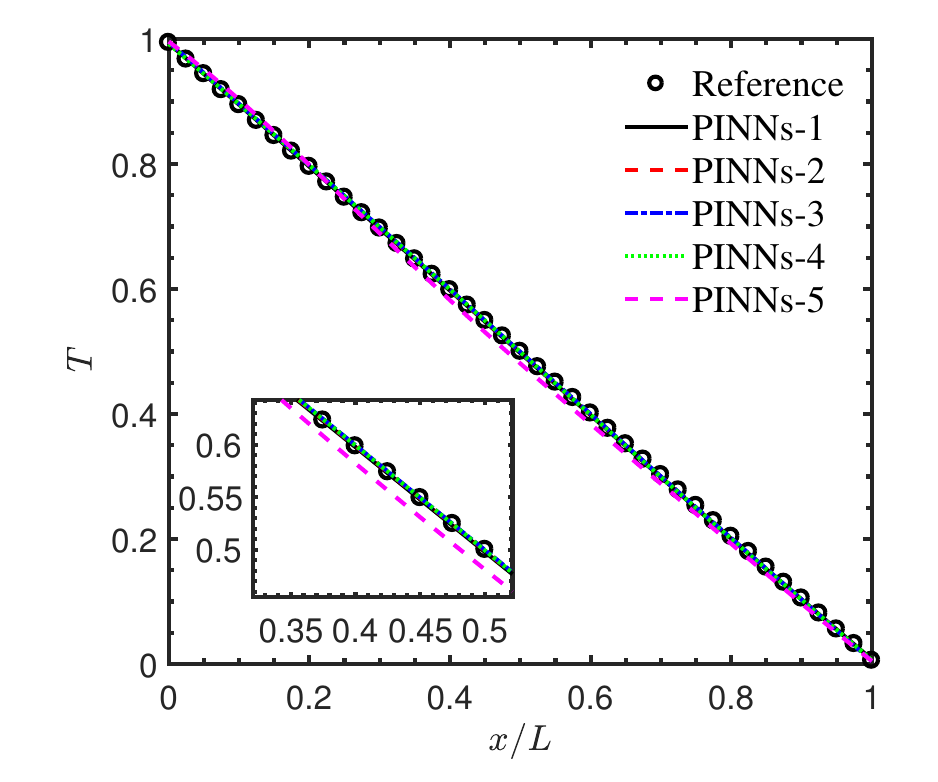}}
\subfigure[]{\label{fig:1d_methods_b}
\includegraphics[width=0.3\textwidth]{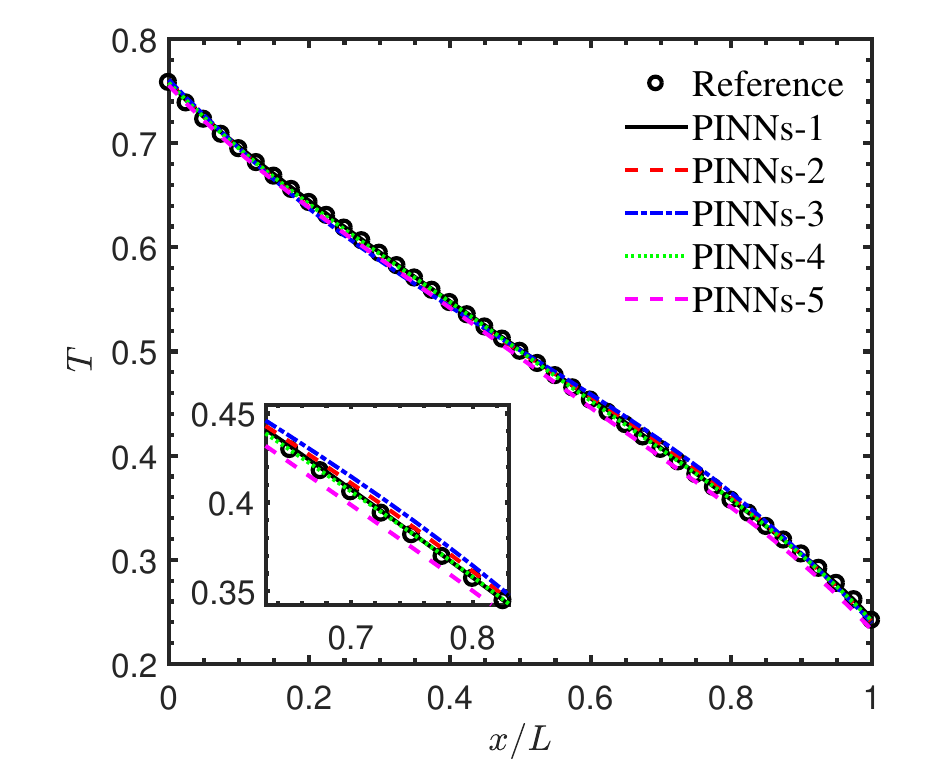}}
\subfigure[]{\label{fig:1d_methods_c}
\includegraphics[width=0.3\textwidth]{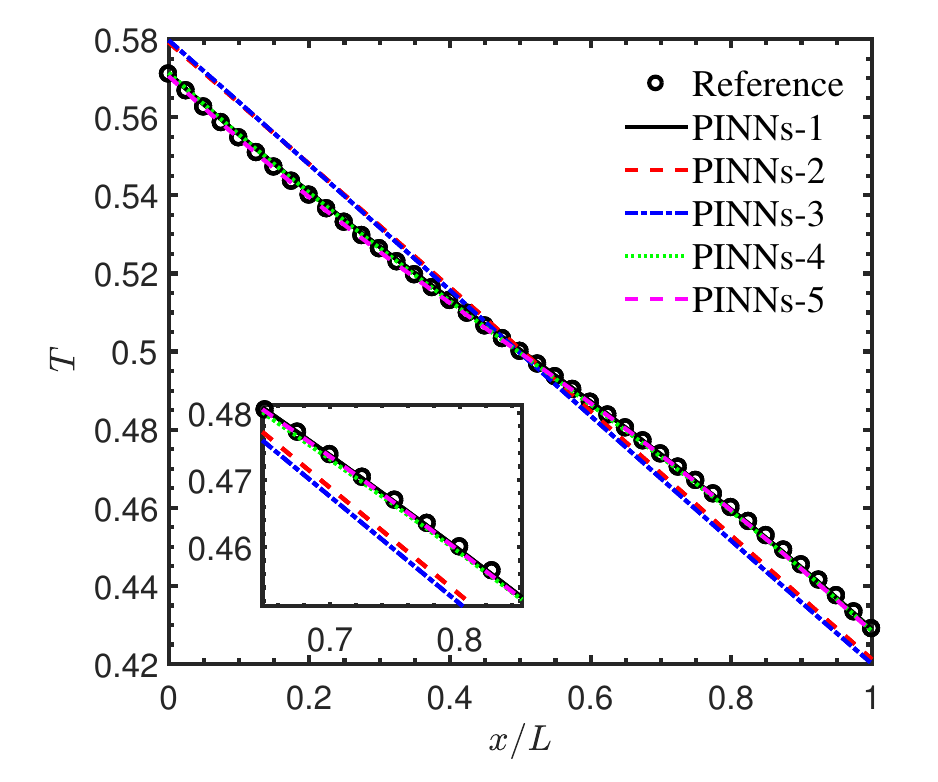}}
\caption{Quasi-1D film heat conduction: Effects of different sampling approaches in solid angular space on the predicted accuracy at (a) $\rm{Kn}=0.01$, (b) $\rm{Kn} = 1 $, and (c) $\rm{Kn} =10$. Symbols: Reference solution \cite{zhang2023acceleration}; PINNs-1: Two-step sampling method: Random sampling in both spatial and solid angular domain; PINNs-2: uniform discrete points in both spatial and solid angular domain; PINNs-3: uniform discrete points in spatial domain and GL quadrature points in angular domain; PINNs-4: Two-step sampling method: uniform discrete points in spatial domain and random sampling in solid angular space; PINNs-5: random sampling in spatial-angular domain.}
\label{FIG:1d_methods}
\end{figure}

We utilize these five sampling methods in PINNs to model the heat conduction in the 1D film at $\rm{Kn} = 0.01$, 1, and 10.  The results and the computational errors are present in Fig.~\ref{FIG:1d_methods} and Table \ref{tab:1d_comp}, respectively.  As we can see, (1) the PINNs with the two-step sampling approach, i.e., PINNs-1 and PINNs-4, are able to provide the most accurate results for the heat conduction from diffusive to ballistic regimes; (2) the method in \cite{li2021physics}, i.e., PINNs-3, and PINNs-2, are able to achieve good accuracy for Kn $=0.01$ and 1, but failed to capture the non-Fourier's behavior accurately at Kn $= 10$ (e.g., the temperature jumps at the boundaries), as shown in Fig. \ref{fig:1d_methods_c}; and (3) the computational errors for PINNs-5 increases with the decreasing Kn. As discussed in Sec. \ref{sec:two_step}, here we only solve the BTE at one point in the solid angular space at each spatial point, leading to poor accuracy for the computation of $e^{eq}$.  At the ballistic scale, i.e., cases with large Kn, we have very few collisions for the phonon, and the inaccurate estimation of $e^{eq}$ has little effect on the predicted accuracy. However, for cases with small Kn, we have frequent collisions of the phonon, and the inaccurate prediction of $e^{eq}$ results in poor accuracy for solving the BTE. 

\begin{table}[H] 
\caption{\label{tab:1d_comp}
Quasi-1D film heat conduction: Relative $L_2$ errors of the temperatures between PINNs and reference solution with different training points. PINNs-1 and PINNs-4: two-step sampling approach proposed in the current study; PINNs-2: uniform grids in both domains; PINNs-3: sampling approach used in \cite{lou2021physics, li2021physics}; PINNs-5: sampling approach used in \cite{jin2023asymptotic}. 
}
\newcolumntype{C}{>{\centering\arraybackslash}X}
\begin{tabularx}
{\textwidth}{m{3cm} XXXXX}
\toprule
 - & PINNs-1	& PINNs-2 & PINNs-3 & PINNs-4 & PINNs-5\\
\midrule
$E_T(\rm{Kn} = 0.01)$		& \bf{0.402}\% & 0.383\%& 0.370\% & \bf{0.390}\% & 2.112\%\\
$E_T(\rm{Kn} = 1)$		& \bf{0.299}\% & 1.005\%& 0.678\% & \bf{0.214}\% & 1.224\%\\
$E_T(\rm{Kn} = 10)$		& \bf{0.092}\% &1.280\% & 1.076\%& \bf{0.065}\% & 0.115\%\\
\bottomrule
\end{tabularx}
\end{table}

\subsection{Unsteady quasi-1D film heat conduction with space varying Knudsen number}

\begin{figure}[H]
    \centering
    \includegraphics[width=0.5\textwidth]{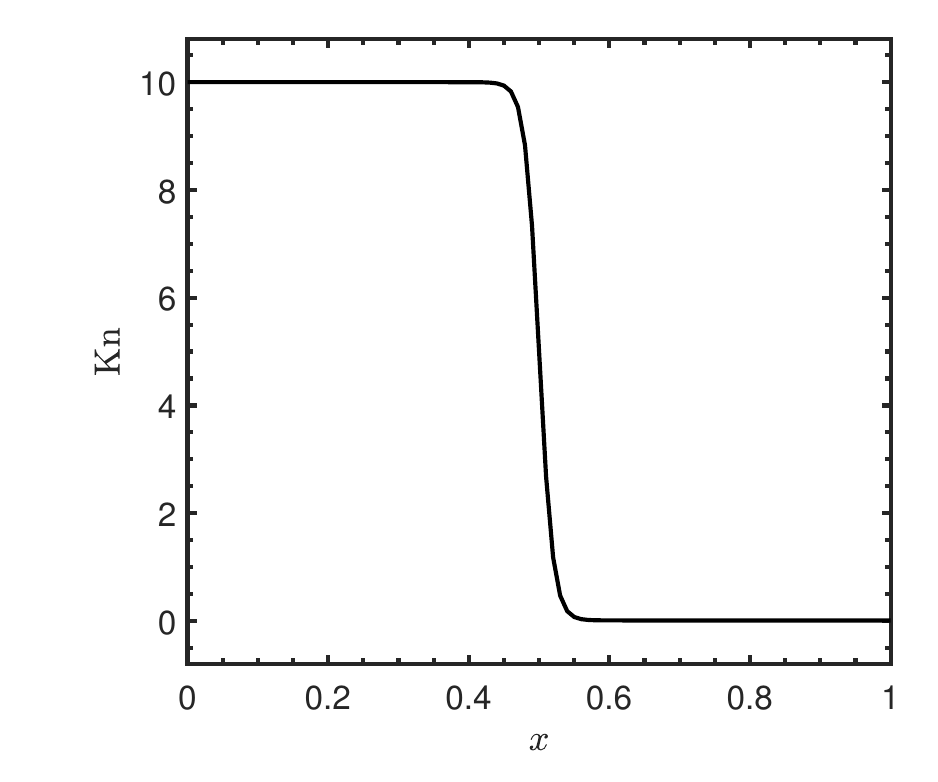}
    \caption{
    The varying Knudsen number within the film.
    }
    \label{fig:tau}
\end{figure}

In this section, we apply the MC-PINNs to model the time-dependent heat conduction in a quasi-1D film with a varying Knudsen number, which is defined as follows:
\begin{equation}
    {\rm{Kn}}(x) = \frac{{\rm{Kn}}_{\max}+{\rm{Kn}}_{\min}}{2} - \frac{{\rm{Kn}}_{\max}-{\rm{Kn}}_{\min}}{2}\tanh\left({\frac{x-x_c}{2d}}\right),
\end{equation}
where $\rm{Kn}_{\max}$ and $\rm{Kn}_{\min}$ represent the maximum and minimum Knudsen number, respectively; $x_c = L/2$ is the center of the film with $L$ denoting the thickness of the film, and $d = 0.01$ characterizes the thickness of the transition layer. In the present case, we set $\rm{Kn}_{\max} = 10$ and $\rm{Kn}_{\min} = 0.01$.
The distribution of the Knudsen number within the film is shown in Fig.~\ref{fig:tau}. It can be seen that we have a significant change in $\rm{Kn}$ near the center of the film. The heat conduction at the left half is at the ballistic regime, and is at the diffusive regime on the right half of the film.

The computational domain as well as the boundary conditions are kept the same as in Sec. \ref{sec:1d}.  In addition, we set $T(t = 0, x ) = (T_h + T_c)/2$ in the entire domain. The time domain here is set as $t \in [0, 10]$. In the computations, we sample $B_{t,x} = 100$ random points in the temporal-spatial domain and $B_s = 16$ random points in the solid angular domain. Further, $B_{t, x_w,\bm{s}}=100$ random points are generated for boundary conditions and $B_{t_0, x,\bm{s}}=100$ random points for initial conditions.

\begin{figure}[H]
\centering
\subfigure[]{\includegraphics[width=0.45\textwidth]{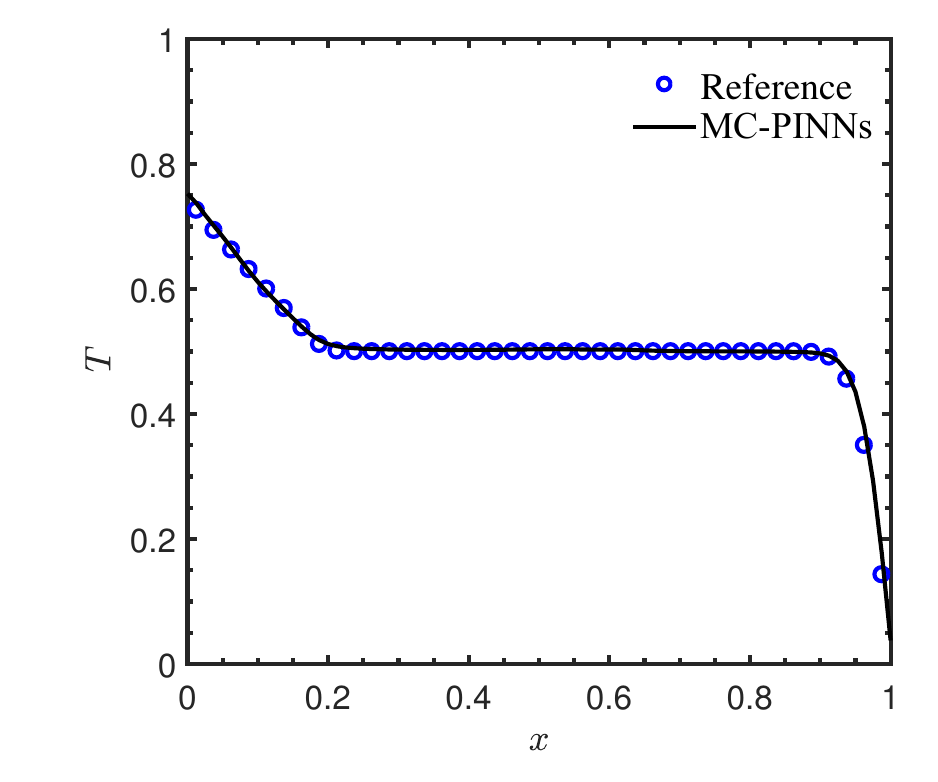}}
\subfigure[]{\includegraphics[width=0.45\textwidth]{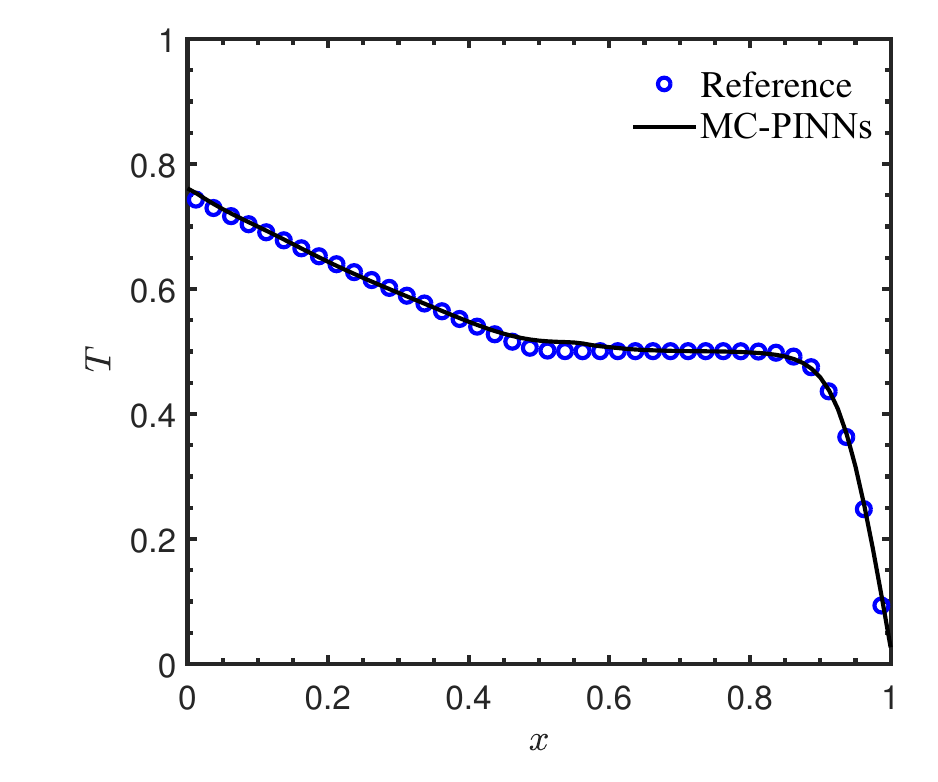}}
\subfigure[]{\includegraphics[width=0.45\textwidth]{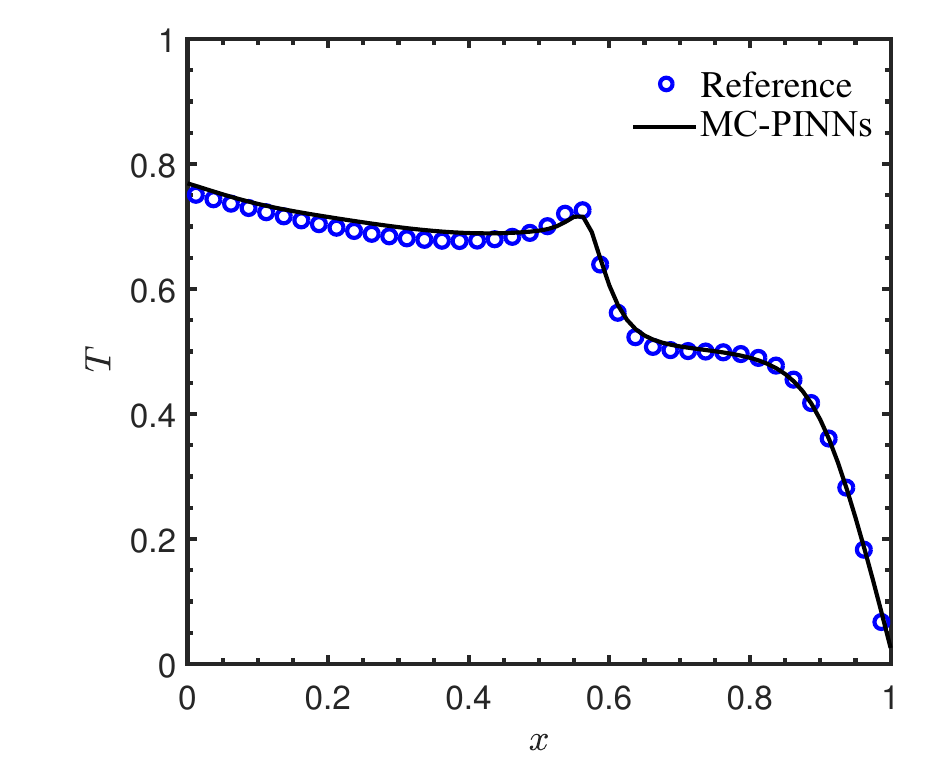}}
\subfigure[]{\includegraphics[width=0.45\textwidth]{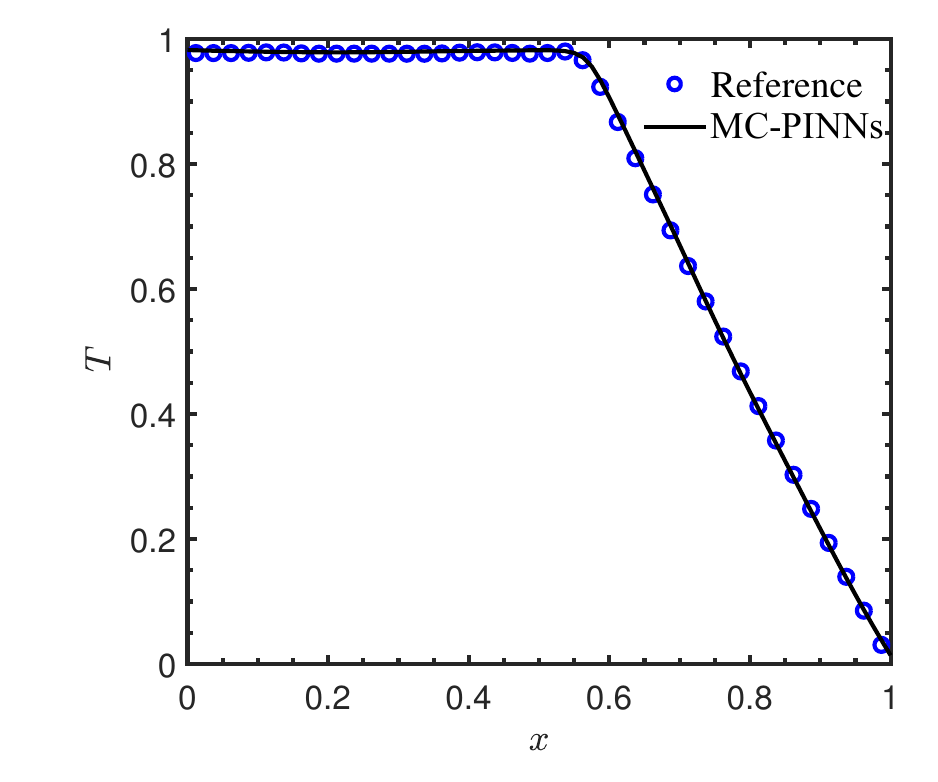}}
\caption{Unsteady quasi-1D film heat conduction: The predicted temperature at different times with space varying Knudsen number. (a) $t=0.2$, (b) $t=0.5$, (c) $t=1$, (d) $t=10$.
Symbols: Reference solution \cite{guo2016discrete}; Black solid line: MC-PINNs.}
\label{FIG:1D_t}
\end{figure}

Figure \ref{FIG:1D_t} presents the results obtained from the MC-PINNs at four representative times: $t = 0.2$, 0.5, 1, and 10. As shown, the results from the MC-PINNs show little differences with the reference solutions~\cite{guo2016discrete} at different times, demonstrating the good accuracy of the MC-PINNs for modeling multiscale heat conduction.

We would like to point out that 100 Gauss-Legendre quadrature points are employed in the solid angular space in the numerical method to accurately capture the non-Fourier's behavior~\cite{guo2016discrete}. In MC-PINNs,  we only use 16 points in the solid angular domain at each training step, and it is capable of providing good results for heat conduction ranging from diffusive to ballistic scales. The above results indicate the great flexibility of the MC-PINNs. We do not need to carefully design the discretization of the angular domain, which is a long-standing challenge in conventional numerical methods for simulating multiscale heat conduction problems.


\subsection{Quasi-2D square heat conduction}
\label{sec:2d}
\begin{figure}[H]
    \centering
    \includegraphics[width=0.6\textwidth]{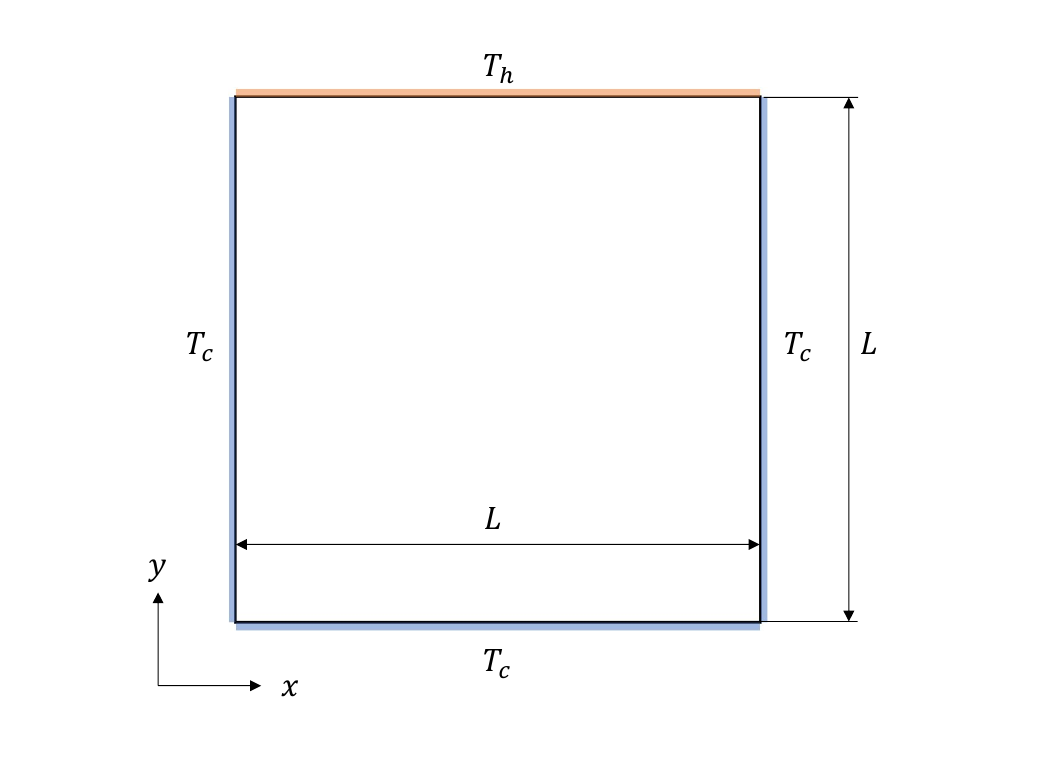}
    \caption{
    Schematic of the quasi-2D heat conduction in a square domain. $L$ is the characteristic length, $T_h$ is the upper plate dimensional temperature and $T_c$ is the dimensional temperature of the remaining plates.
    }\label{fig:square}
\end{figure}

We now consider a quasi-2D heat conduction problem in a square domain with the length $L = 1$. The temperature is set to $T_h = 1$ at the top boundary and $T_c = 0$ at the remaining boundaries, as shown in Fig.~\ref{fig:square}. 


In this section, we employ the MC-PINNs to model the heat conduction in the 2D square domain at $\rm{Kn}=0.1, 1$ and $10$. 
In our computations, we first randomly sample $B_{\bm{x}}=100$ points in the spatial domain, and then select $B_{\bm{s}}=64$ random points in the solid angular domain $(s_x, s_y)$. As for boundary conditions, we randomly select $B_{\bm{x,s}} = 100$ in the spatial-angular domain at each boundary to implement the thermalization boundary condition defined in Eq.~\eqref{eq:bc}.

\begin{figure}[H]
    \centering
    \includegraphics[width=1.0\textwidth]{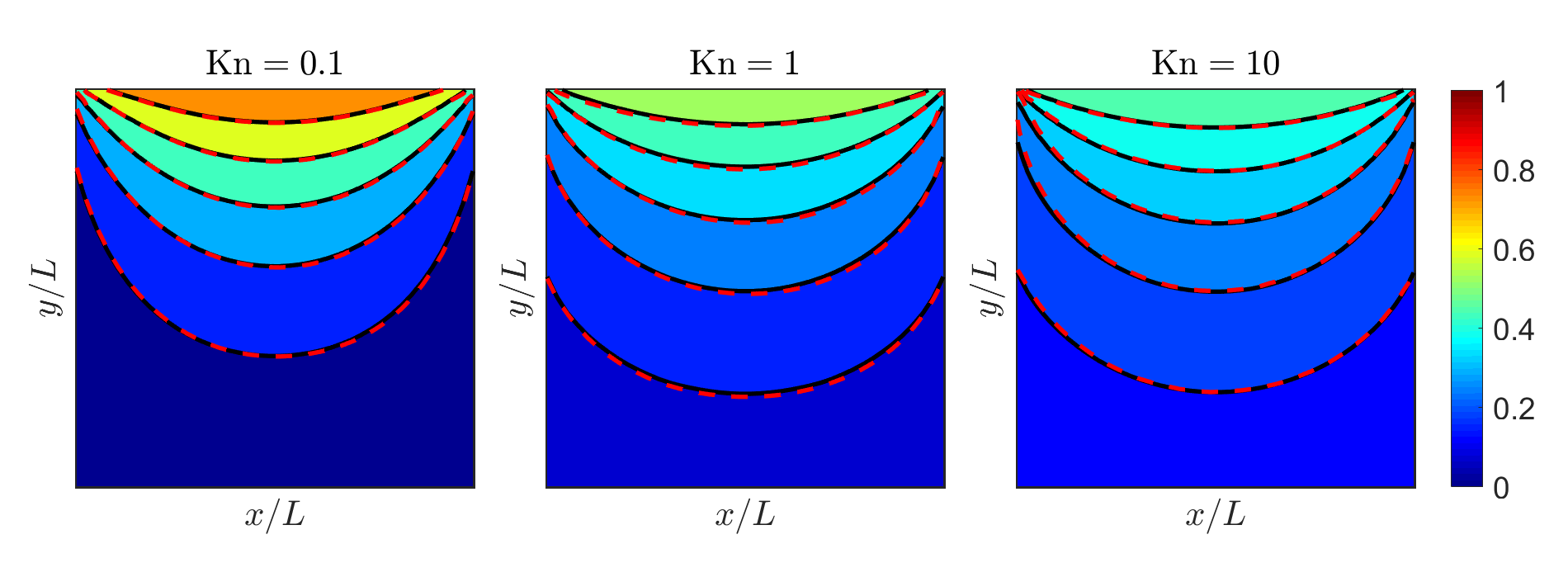}
    \caption{
    Quasi-2D heat conduction: Distribution of the temperature at (a) $\rm{Kn} = 0.1$, (b) $\rm{Kn} = 1$ and (c) $\rm{Kn} = 10$. Colored background with black solid line: reference solution \cite{zhang2023acceleration}; Red dashed line: the MC-PINNs.
    }\label{FIG:cavity contour}
\end{figure}

\begin{figure}[H]
    \centering
    \includegraphics[width=0.6\textwidth]{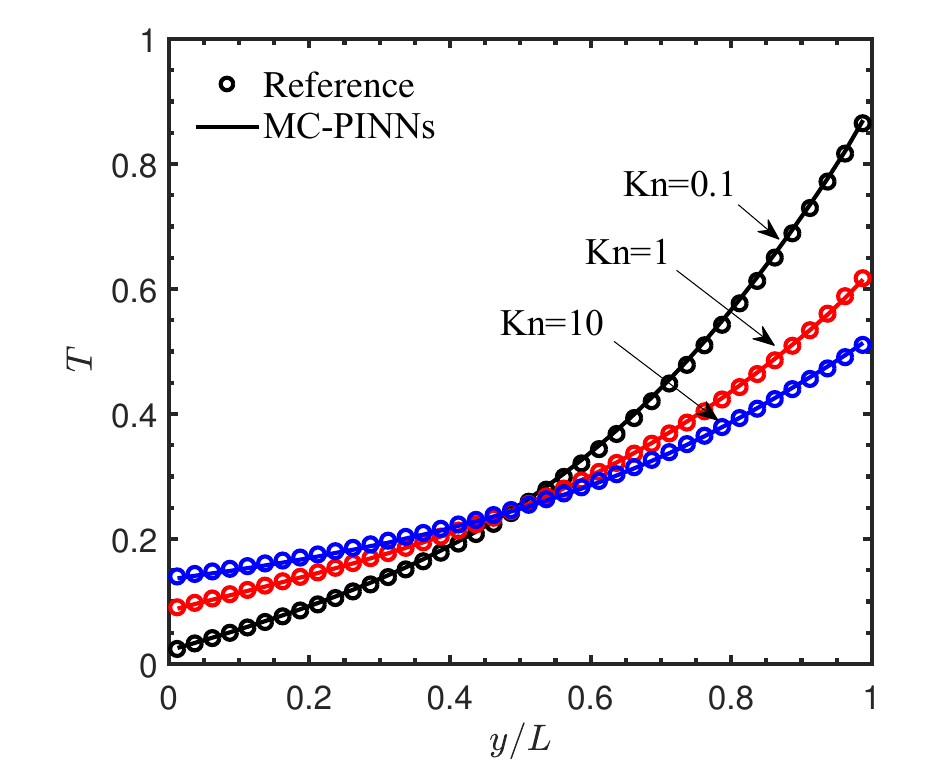}
    \caption{
    Quasi-2D heat conduction: Temperature along the center vertical line at different $\rm{Kn}$. Symbols: Reference solution\cite{zhang2023acceleration}; Solid line: MC-PINNs.}
    \label{fig:square1}
\end{figure}

The results of MC-PINNs are shown in Figs.~\ref{FIG:cavity contour} and \ref{fig:square1}. We can see that the predicted temperatures from the MC-PINNs are in good agreement with the reference solutions \cite{zhang2023acceleration} in all test cases. We also observe in Figs.~\ref{FIG:cavity contour} and \ref{fig:square1} there are significant temperature jumps at the left/right boundaries. In addition,  the temperature jump becomes more pronounced with the increase of $\rm{Kn}$. This phenomenon is reasonable since the non-equilibrium effects become more pronounced in the transition and ballistic regimes, leading to distinct temperature discontinuities at the walls. 


We further compute the relative $L_2$ error between the MC-PINNs and the reference solutions \cite{zhang2023acceleration} at different $\rm{Kn}$ to demonstrate the accuracy of the present method in quasi-2D problems. As shown in Table~\ref{tab3}, the relative $L_2$ errors for the three test cases are smaller than $2\%$, which again confirms the good accuracy of MC-PINNs for modeling multiscale heat conduction.

\begin{table}[H] 
\caption{Quasi-2D heat conduction problem: Relative $L_2$ errors between MC-PINNs and reference solution at different $\rm{Kn}$.\label{tab3}}
\newcolumntype{C}{>{\centering\arraybackslash}X}
\begin{tabularx}
{\textwidth}{CCCC}
\toprule
 - & $\rm{Kn}=0.1$ & $\rm{Kn}=1$ & $\rm{Kn}=10$\\
\midrule
$E_T$			& 1.55\% & 0.62\%& 1.21\% \\
\bottomrule
\end{tabularx}
\end{table}

\subsection{3D cuboid phonon transport}\label{sec:3D}

\begin{figure}[H]
    \centering
    \includegraphics[width=0.6\textwidth]{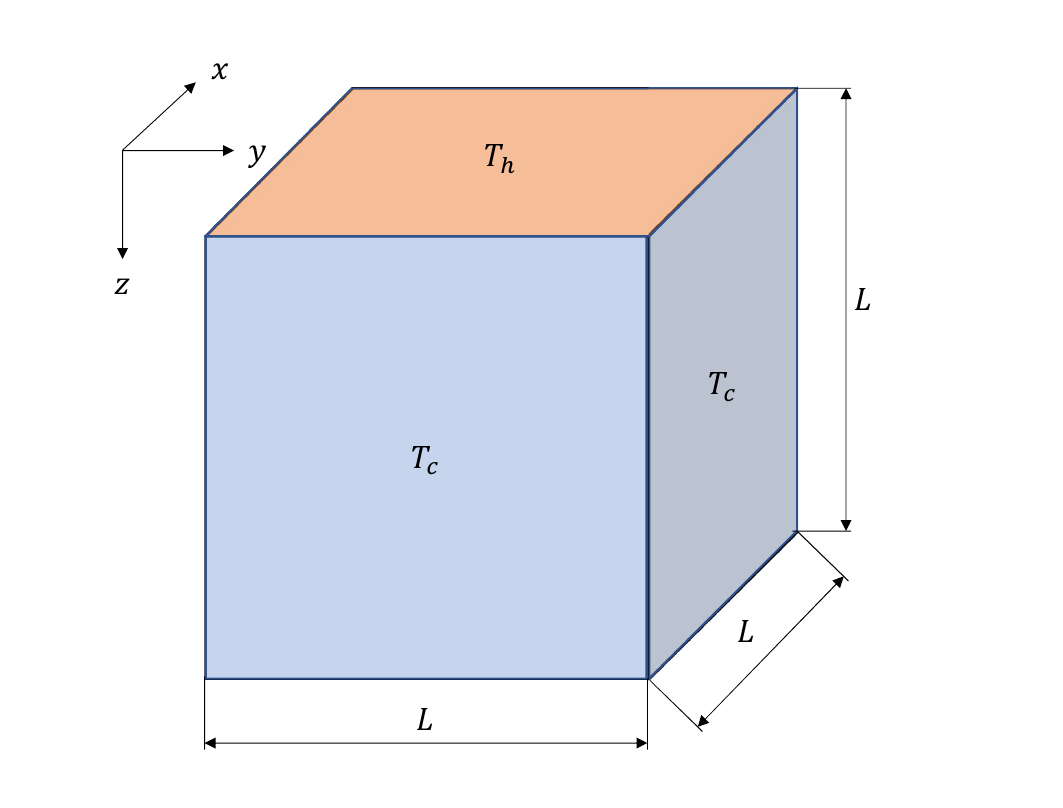}
    \caption{
    Schematic of the three-dimensional heat conduction in a cuboid. }
    \label{fig:cuboid}
\end{figure}

In this section, we proceed to consider the phonon transport in a 3D cuboid. The temperature is set as $T_h$ at the top boundary and $T_c$ at the remaining boundaries, as shown in Fig.~\ref{fig:cuboid}. Specifically, we set $T_h = 1$,  $T_c = 0$, and $L = 1$. 

In the MC-PINNs, the two-step sampling strategy is applied for generating the training points for the loss of residual with $B_{\bm{x}}=100$ in the spatial domain and  $B_{\bm{s}}=64$ in the solid angular domain $(s_x, s_y, s_z)$. For the boundary conditions, we randomly select $B_{\bm{x,s}} = 100$ in the spatial-angular domain for each wall at each training step. The boundary condition implemented here is defined in Eq.~\eqref{eq:bc}, which is the same as the previous two cases in Secs. \ref{sec:1d} and \ref{sec:2d}. As for the reference solution, we employ the implicit kinetic scheme to solve the stationary phonon BTE developed in \cite{zhang2023acceleration}. We note that the numerical method in \cite{zhang2023acceleration} employs a macroscopic equation to accelerate convergence in the diffusive regime. Additionally, it incorporates a memory reduction technique to reduce memory usage in computations. We note that $40 \times 40$ and $80 \times 80$ discrete points in the angular space are employed in the numerical method for $\rm{Kn}=0.1, 1$ and $\rm{Kn}=10$, respectively, to ensure the convergence.  More details for the discretization of the spatial space in the numerical method are present in \ref{sec:appendix_b}.



\begin{figure}[H]
    \centering
    \includegraphics[width=1.0\textwidth]{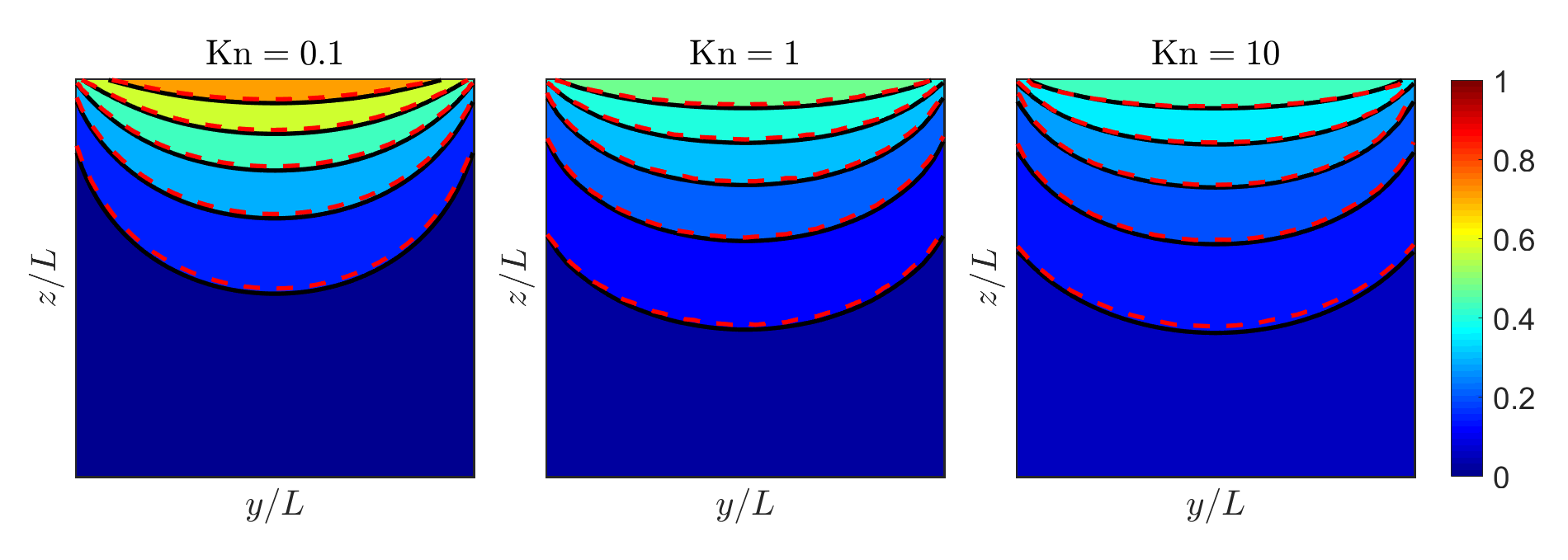}
    \caption{3D heat conduction problem: Predicted temperature from MC-PINNs at $x = 0.5$ for different $\rm{Kn}$. Colored background with black solid line: reference solution; Red dashed line: the MC-PINNs.
    }\label{FIG:3d}
\end{figure}

\begin{figure}[H]
    \centering
    \includegraphics[width=0.5\textwidth]{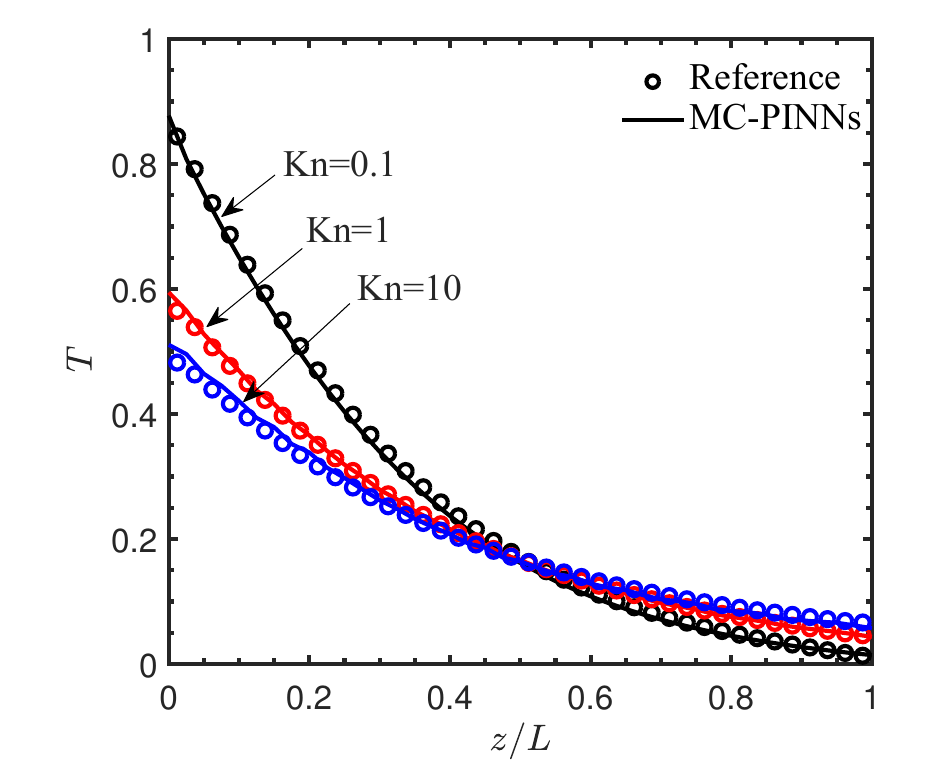}
    \caption{
    3D heat conduction problem: Predicted temperature from MC-PINNs at $x = 0.5$ and $y = 0.5$. Symbols: Reference solution~\cite{zhang2023acceleration}; Solid line: MC-PINNs. }
    \label{fig:3D_1D}
\end{figure}

The predicted temperature from MC-PINNs in each case at $x = 0.5$ are present in Figs. \ref{FIG:3d} and \ref{fig:3D_1D}.  Again, good agreements are observed for all test cases between the MC-PINNs and the reference solutions. Similar to in the 2D cases, there are significant temperature jumps at the boundaries when $\rm{Kn} \ge 1$, which have been discussed in Sec. \ref{sec:2d} and will not present here for simplicity. We then illustrate the relative $L_2$ errors between the temperature from the MC-PINNs and the reference solutions at different $\rm{Kn}$  in Table~\ref{tab4}. The results further confirm the capability of the MC-PINNs for 3D multiscale heat conduction problems.


\begin{table}[H] 
\caption{3D heat conduction problem: Relative $L_2$ errors between the temperature from MC-PINNs and reference solution at different $\rm{Kn}$.\label{tab4}}
\newcolumntype{C}{>{\centering\arraybackslash}X}
\begin{tabularx}
{\textwidth}{CCCC}
\toprule
 - & $\rm{Kn}=0.1$ & $\rm{Kn}=1$ & $\rm{Kn}=10$\\
\midrule
$E_T$			& 2.49\% & 1.29\%& 1.91\% \\
\bottomrule
\end{tabularx}
\end{table}


As mentioned, the method in \cite{zhang2023acceleration} is one of the most efficient approaches for solving stationary BTE for all Knudsen numbers. We now perform a comparison of the computational and the memory cost between the MC-PINNs and the numerical method \cite{zhang2023acceleration} that is used to generate the reference solution here. Specifically, we compare the computational time and the memory usage using two representative cases, i.e., Kn $=0.1$ and 10.

We first discuss the computational time for MC-PINNs. As illustrated in Table \ref{tab:comparison1}, the computational time of MC-PINNs trained from scratch (second column in Table \ref{tab:comparison1}) is about 7 and 1.3 times of that in the numerical method \cite{zhang2023acceleration} for Kn $ = 0.1$ and 10, respectively.  In the deep learning community, the transfer learning technique is a widely used approach to speed up the training of DNNs. Here we can also use it to speedup the training of MC-PINNs. Specifically, we first train the MC-PINNs from scratch for Kn $= 1$, and then save the hyperparameters, i.e., weights and biases, in order to reuse them in other cases. In particular, we initialize the MC-PINNs for Kn $=0.1$ and 10 using the hyperparameters from the pre-trained MC-PINNs for Kn $= 1$ and further train them for an additional number of steps until  convergence. The relative $L_2$ errors for these two cases with transfer learning are $2.40\%$ and $2.60\%$, which are similar to the cases without transfer learning.  Also, as we can see in Table \ref{tab:comparison1}, we can achieve about 3-6 times speedup with transfer learning, and the computational times for these two cases are now comparable to the numerical method in \cite{zhang2023acceleration}.

Note that we run the numerical method and MC-PINNs on the CPU and GPU, respectively. In particular, all the CPU computations are conducted on a single Intel Xeon Platinum 9242 processor, and the GPU computations are performed using an NVIDIA GeForce RTX 3090. We remark that the comparison present in Table \ref{tab:comparison1} is not fair since we can also accelerate the numerical method by developing GPU codes and running the tests on GPUs. However, developing GPU codes for the numerical method is more challenging than MC-PINNs and is beyond the scope of the current study. 
We show the results here to provide a preliminary reference on the computational cost of MC-PINNs and one of the state-of-the-art numerical methods in this field. Therefore, we only present the computational time for the numerical method on CPUs. 

\begin{table}[H]
\caption{
Computation time for the 3D heat conduction problem at $\rm{Kn}=0.1$ and 10. First column:  Numerical method on a single Intel Xeon Platinum 9242 processor. Second and third columns:  MC-PINNs w/o and w/ transfer learning (TL). The MC-PINNs are run on one NVIDIA GeForce RTX 3090.
}
\label{tab:comparison1}
\centering
\begin{tabular}{c|c|c|c}
\toprule
{$\rm{Kn}$} & Numerical method & MC-PINNs (w/o TL) & MC-PINNs (w/ TL) \\ \hline

0.1   &  \bf{458}s    & 3386s    & 870s\\ \hline
10    &  2500s    & 3101s   & \bf{445}s \\ 
\bottomrule
\end{tabular}
\end{table}



We now discuss the memory usage for MC-PINNs and the numerical method,  which is present in Table \ref{tab:comparison2}. Note that the MC-PINNs with and without transfer learning have the same memory cost. We therefore only illustrate the memory usage in the MC-PINNs without transfer learning in Table \ref{tab:comparison2}. In the MC-PINNs, we employ the same number of training points for Kn $= 0.1$ and 10, and thus the memory usage is almost the same for these two cases. In the numerical method, the memory cost is more expensive for Kn  $=10$ than for Kn $ = 0.1$ since we have more discrete points in the angular domain for the former case. Also, the MC-PINN requires about $66\%$ and $16\%$ of the memory cost in the numerical method \cite{zhang2023acceleration} for Kn $=0.1$ and 10, respectively. It is worth mentioning that the problem we consider here is relatively small. For real-world applications, we may have more than 40 meshes in each direction of the spatial domain in numerical methods. For example, if we increase the number of meshes in each direction from 40 to 400 and keep the same discretization in the solid angular space for Kn $= 10$, the memory usage will increase to about 1,500GB, which is quite challenging to handle. However, in the MC-PINNs, we can use the same setup as used in the present case, suggesting that the memory usage can be the same, i.e., which is about 6,000 times less than in the numerical method \cite{zhang2023acceleration}.

\begin{table}[H]
\caption{Memory usage for the 3D heat conduction problem at $\rm{Kn}=0.1$ and $\rm{Kn}=10$.}
\label{tab:comparison2}
\centering
\begin{tabular}{c|c|c}
\toprule
 & Kn $= 0.1$ & Kn $ = 10$\\  \hline
Numerical method  & 381MB       & 1562MB\\ \hline
MC-PINNs    & {\bf 252MB}       & {\bf 257MB} \\ \bottomrule
\end{tabular}
\end{table}

We would like to briefly summarize our main findings here: (1) the two-step sampling approach enables the MC-PINNs to use much less memory compared to the numerical method \cite{zhang2023acceleration}, especially for large-scale heat conduction problems spanning the diffusive and ballistic regimes; and (2) the computational time of MC-PINNs (GPU time) with transfer learning is comparable to the numerical method \cite{zhang2023acceleration} (CPU time) for the specific problem considered here. 


\section{Summary}
\label{sec:summary}
In this study, we develop  Monte Carlo physics-informed neural networks (MC-PINNs) for solving the phonon Boltzmann transport equation (BTE) to model the multiscale heat conduction in solid materials.  Specifically, we utilize a deep neural network to approximate the energy density function in BTE and encode the BTE as well as the boundary/initial conditions via the automatic differentiation, in MC-PINNs. In addition, we propose a novel two-step sampling approach to address the issues like ``curse of dimensionality'' and the inaccurate estimate of the equilibrium distribution functions in the existing sampling strategies that are widely used in PINNs. In particular, we first randomly sample a certain number of points in the temporal-spatial domains (Step I), and then we draw another number of random points in the solid angular domain (Step II). The residual points at each training step in MC-PINNs are constructed using the tensor product given the points generated at the above two steps.  We conduct a series of numerical experiments to validate the accuracy of the MC-PINNs, including the steady quasi-1D/quasi-2D/3D as well as unsteady quasi-1D heat conduction problems spanning the diffusive and ballistic regimes. The results show the MC-PINNs is capable of achieving good accuracy for the multiscale heat conduction from 1D to 3D.  Furthermore, the present method requires quite a small number of  points in the solid angular domain than in the deterministic methods for solving the BTE, especially for scenarios in the ballistic regimes. For instance, we  employ only 64 points in the solid angular domain at each training step in MC-PINNs for Kn ranging from 0.1 to 10 in the 3D case. In contrast, we have $80 \times 80$ discrete points in the solid angular domain of the numerical method to obtain converged results. The memory cost in the MC-PINNs is about 1/6 of that in the numerical method. Further, we are able to achieve similar computational efficiency for MC-PINNs compared to the numerical method in the 3D test case as we utilize the transfer learning technique in the MC-PINNs.

We would like to discuss that the MC-PINNs developed in this work is mesh-free in the temporal-spatial-angular domain, which is promising in real-world applications with complicated domains, for instance, the design of microdevices. In addition, the MC-PINNs is capable of providing good accuracy for heat conduction ranging from diffusive to ballistic with only a small number of random discrete points in the solid angular domain at each training step. It can save lots of effort in numerical methods since the optimal discretization of the solid angular space at different regimes is generally different and should be carefully designed to ensure convergence in these methods. Also, the memory efficacy and the flexibility of treatment for the solid angular space in the present method make it a promising tool for modeling multiscale heat conduction in large-scale problems. Finally, we note that the MC-PINNs with the two-step sampling approach can be seamlessly extended to solve kinetic equations used in other disciplines, e.g., the Boltzmann-BGK model for multiscale flows, the radiation transfer equation for multiscale heat transfer, etc. We will leave these interesting topics as future studies.


\section*{Acknowledgements}
Q. L., X. M., and Z. G. acknowledge the support of the National Natural Science Foundation of China (No. 12201229) and the Interdisciplinary Research Program of HUST (No. 2024JCYJ003 and 2023JCYJ002).
C. Z. was supported by the National Natural Science Foundation of China (No. 12147122). X. M. would also like to acknowledge the support of Xiaomi Young Talents Program. The authors acknowledge the  Beijng PARATERA Tech CO., Ltd. for the HPC resources.

\appendix

\section{Gradients unbiasedness and convergence for the two-step sampling method in MC-PINNs}\label{sec:appendix_a}


In this section, we follow \cite{garrigos2023handbook} to conduct a theoretical analysis of the unbiasedness of the gradient and the convergence of the two-sampling method in MC-PINNs.  Note that the sampling approach to compute the losses for the boundary and/or initial conditions are the same as the widely used minibatch training in the deep learning community, we thus only focus on the analysis of the residual loss in what follows.


We begin with the optimization trajectory of the gradient descent:
\begin{equation}\label{eq:trajectory}
    \theta_{n+1}=\theta_n-\gamma\nabla f(\theta_n),
\end{equation}
where 
$\theta_n$ denotes the parameters at $n$-th training step, $\gamma$ is the learning rate, and $f(\theta_n)$ is the loss function for the residual of the governing equation with $f: \mathbb{R}^d\rightarrow\mathbb{R}$.
Assuming that the full batch is constructed using sufficient points from the temporal-spatial and angular domains via a tensor product. The full-batch gradient can then be expressed as:
\begin{equation}
\nabla f(\theta)=\frac{1}{N_{t,\bm{x}}N_{\bm{s}}}\sum_{i=1}^{N_{t,\bm{x}}}\sum_{j=1}^{N_{\bm{s}}}\nabla f_{i,j}(\theta),
\end{equation}
where $N_{t,\bm{x}}$ and $N_{\bm{s}}$ represent the numbers of full batch points in the temporal-spatial and angular domains, respectively.


In the two-step sampling approach, we randomly select two subsets from the full batches in the temporal-spatial and angular domains, respectively. The corresponding optimization trajectory for the two-step sampling method can be expressed as:
\begin{equation}\label{eq:trajectory1}
    \theta_{n+1}=\theta_n-\gamma\nabla f_{B_{t, \bm{x}},B_{\bm{s}}}(\theta_n),
\end{equation}
where
\begin{equation}\label{eq:sgd_gradient}
    \nabla f_{B_{t, \bm{x}},B_{\bm{s}}}(\theta)=\frac{1}{B_{t, \bm{x}}B_{\bm{s}} }\sum_{i=1}^{B_{t,\bm{x}}}\sum_{j=1}^{B_{\bm{s}}} \nabla f_{i,j}(\theta_n).
\end{equation}
Here, $B_{t, \bm{x}}$ and $B_{\bm{s}}$ denote the number of points in subsets in the temporal-spatial and angular space domains, respectively.
The expectation of the gradient in Eq. \eqref{eq:sgd_gradient} can be derived as follows:

\begin{equation}
\begin{aligned}\label{eq:E_batch}
    \mathbb{E}[\nabla f_{B_{t, \bm{x}},B_{\bm{s}}}(\theta)]&=\mathbb{E}\left[\frac{1}{B_{t, \bm{x}}B_{\bm{s}} }\sum_{i=1}^{B_{t,\bm{x}}}\sum_{j=1}^{B_{\bm{s}}} \nabla f_{i,j}(\theta)\right]\\
    &=\frac{1}{B_{t, \bm{x}}B_{\bm{s}}}\sum_{i=1}^{B_{t,\bm{x}}}\sum_{j=1}^{B_{\bm{s}}} \mathbb{E}[\nabla f_{i,j}(\theta)] \\
    & = \frac{1}{B_{t, \bm{x}}B_{\bm{s}}}\sum_{i=1}^{B_{t,\bm{x}}}\sum_{j=1}^{B_{\bm{s}}} \nabla f(\theta)\\
    & = \nabla f(\theta),
\end{aligned}
\end{equation}
which demonstrates that the gradient obtained by the two-step sampling method is unbiased.


We now conduct the analysis on the convergence of the two-step sampling method in MC-PINNs.
Since $f$ is continuous and the number of full batch training points is finite, we can define the gradient noise in the two-step sampling method as follows according to \cite{garrigos2023handbook} :
\begin{equation}
    \sigma^*=\inf_{\theta^*\in \mathop{\arg\min} f}\mathbb{V}[\nabla f_{B_{t,\bm{x}},B_{\bm{s}}}(\theta^*)],
\end{equation}
where $\theta^*\in \mathop{\arg\min} f$ denotes the NN parameters when $f$ reaches the global minimum. 
Based on the trajectory in Eq.~\eqref{eq:trajectory1}, we can obtain the following equation:
\begin{equation}
\begin{aligned}
     \Vert \theta_{n+1} - \theta^*\Vert^2 &=  \Vert \theta_{n} -\gamma\nabla f_{B_{t, \bm{x}},B_{\bm{s}}}(\theta) - \theta^*\Vert^2\\
    &=\Vert\theta_n-\theta^*\Vert^2 - 2\gamma\langle \theta_n-\theta^*,\nabla f_{B_{t,\bm{x}},B_{\bm{s}}}(\theta_n)\rangle + \gamma^2\Vert \nabla f_{B_{t,\bm{x}},B_{\bm{s}}} (\theta_n)\Vert^2.
\end{aligned}
\end{equation}
Taking the expectation conditioned on $\theta_n$ and utilizing the unbiasedness of $\nabla f_{B_{t, \bm{x}},B_{\bm{s}}}$  in Eq.~\eqref{eq:E_batch}, we obtain:
\begin{equation}\label{eq:E_theta}
\begin{aligned}
        \mathbb{E}[\Vert \theta_{n+1} - \theta^*\Vert^2\vert\theta_n]&=\Vert\theta_n-\theta^*\Vert^2 - 2\gamma\langle\theta_n-\theta^*,\nabla f(\theta_n) \rangle + \gamma^2\mathbb{E}[\Vert \nabla f_{B_{t, \bm{x}},B_{\bm{s}}} (\theta_n)\Vert^2\vert \theta_n].
\end{aligned}
\end{equation}
Assuming that $f$ is a convex and differentiable function, we apply Lemma 2.8 in~\cite{garrigos2023handbook}, which states:
\begin{equation}
    f(\theta^*) - f(\theta_n)\geq \langle\theta^* - \theta_n,\nabla f(\theta_n) \rangle,
\end{equation}
we can then obtain the following inequality:
\begin{equation}\label{eq:E_leq1}
    \mathbb{E}[\Vert \theta_{n+1} - \theta^*\Vert^2|\theta_n]\leq \Vert\theta_n-\theta^*\Vert^2 - 2\gamma (f(\theta_n) -f(\theta^*)) + \gamma^2\mathbb{E}[\Vert \nabla f_{B_{t, \bm{x}},B_{\bm{s}}}(\theta_n)\Vert^2|\theta_n].
\end{equation}


We then assume that $f$ in the minibatch is $L_b$-smooth in expectation, which is defined for $\theta_i, \theta_j \in \mathbb{R}^d$ as:
\begin{equation}
    \frac{1}{2L_b}\mathbb{E}[\Vert \nabla f_{B_{t, \bm{x}},B_{\bm{s}}}(\theta_i)-\nabla f_{B_{t, \bm{x}},B_{\bm{s}}}(\theta_j)\Vert^2]\leq f(\theta_i) - f(\theta_j) - \langle \nabla f(\theta_j), \theta_i - \theta_j),
\end{equation}
where $L_b$ is the Lipschitz constant. Then, following the Lemma 6.7 in \cite{garrigos2023handbook}, we can further obtain that:
\begin{equation}
    \mathbb{E}[\Vert\nabla f_{B_{t, \bm{x}},B_{\bm{s}}}(\theta)\Vert^2]\leq 4L_b(f(\theta) - f(\theta^*))+2\sigma^*.
\end{equation}
Consequently, the inequality Eq.~\eqref{eq:E_leq1} can be rewritten as:
\begin{equation}\label{eq:E_leq2}
    \mathbb{E}[\Vert \theta_{n+1} - \theta^*\Vert^2|\theta_n]\leq 
    \Vert\theta_n-\theta^*\Vert^2 + 2\gamma(2\gamma L_b-1)(f(\theta_n)-f(\theta^*))+ 2\gamma^2\sigma^*.
\end{equation}
We now reach the following inequality if we set the learning rate to satisfy $0<\gamma\leq 1/4L_b$~\cite{garrigos2023handbook}:
\begin{equation}
    \mathbb{E}[\Vert \theta_{n+1} - \theta^*\Vert^2|\theta_n]\leq \Vert\theta_n-\theta^*\Vert^2 - \gamma (f(\theta_n)-f(\theta^*))+ 2\gamma^2\sigma^*.
\end{equation}
We then take expectation to obtain the following inequality:
\begin{equation}
    \mathbb{E}[\Vert \theta_{n+1} - \theta^*\Vert^2]\leq \mathbb{E}[\Vert \theta_{n} - \theta^*\Vert^2] - \gamma \mathbb{E}[f(\theta_n) - f(\theta^*)] + 2\gamma^2\sigma^*.
\end{equation}
Summing over $i = 0, 1, ... , n-1$, the above inequality can be written as:
\begin{equation}
        \mathbb{E}[\Vert \theta_n - \theta^* \Vert^2]\leq \Vert \theta_0 - \theta^*\Vert^2 - \gamma \sum_{i=0}^{n-1} \mathbb{E} [f(\theta_i) - f(\theta^*)]
        + 2n\gamma^2\sigma^*.
\end{equation}
Since $\mathbb{E}[\Vert \theta_n - \theta^* \Vert^2]\geq 0$, we now obtain:
\begin{equation}
    \sum_{i=0}^{n-1} \mathbb{E} [f(\theta_i) - f(\theta^*)]\leq \frac{\Vert \theta_0 - \theta^*\Vert^2}{\gamma} + 2n\gamma\sigma^*.
\end{equation}
Defining $\Bar{\theta}_n = \frac{1}{n}\sum_{i=0}^{n-1} \theta_i$ and invoking Jensen's inequality for the convex function $f$ gives:
\begin{equation}
    \mathbb{E} [f(\Bar{\theta}_n) - f(\theta^*)]\leq \frac{\Vert \theta_0 - \theta^*\Vert^2}{\gamma n} + 2\gamma\sigma^*.
\end{equation}
Specifically, when the training step $n$ is fixed and $n \geq 1$, we set $\gamma = \gamma_0/\sqrt{n}$:
\begin{equation}
    \mathbb{E}[f(\Bar{\theta}_n)-f(\theta^*)]\leq \frac{\Vert \theta_0-\theta^*\Vert^2}{\gamma_0 \sqrt{n}}+\frac{2\gamma_0\sigma^*}{\sqrt{n}}=\mathcal{O}(\frac{1}{\sqrt{n}}),
\end{equation}
where $\gamma_0$ is a constant with $0<\gamma_0\leq 1/4L_b$.
The above results demonstrate the convergence of the two-sampling approach as the number of training steps $n$ becomes sufficiently large.

It is important to note that the optimization of PINNs generally involves a non-convex loss landscape. However, as reported in \cite{kawaguchi2016deep},  when the PINNs converge to a local minimum at  $\theta^{**}$, where the Hessian matrix of the loss function at $\theta^{**}$ is positive definite~\cite{hu2024tackling}, the predicted accuracy is in general good.  Further, the convergence to a good local minimum requires appropriate initialization of the neural networks. Empirically, the widely used Xavier's initialization~\cite{glorot2010understanding}, which is also employed in this study, is found to be able to provide good initializations in MC-PINNs.

\section{Details on the computations}
\label{sec:appendix_b}

In all the test cases of Sec.~\ref{sec:results}, the Adam optimizer with a learning rate $10^{-3}$ is employed for the training of PINNs. More details on the architectures of MC-PINNs, etc.,  are present in Table~\ref{app._tab1}.  In addition, 64 points are utilized to compute the equilibrium in each case using the Monte Carlo integration. We note that we have tested the accuracy with more than 64 points, which show little difference from the results presented in this study. Therefore, we employ 64 points for computational efficiency. 

\begin{table}[H]
\caption{Architectures and training steps for MC-PINNs in each case.
}\label{app._tab1}
\newcolumntype{C}{>{\centering\arraybackslash}X}
\begin{tabularx}
{\textwidth}{m{2.5cm}CCC}
\toprule
   & Width $\times$ Depth	& Activation function & Training steps \\
\midrule
Sec. 3.1		& $20\times 4 $ & tanh& 50,000/100,000 \\
Sec. 3.2		& $30\times 4 $&tanh & 100,000 \\
Sec. 3.3		& $30\times 4 $ & tanh& 200,000 \\
Sec. 3.4		& $30\times 5 $ &tanh & 200,000 \\
Appendix C  & $20\times 4 $ & tanh& 50,000          \\
\bottomrule
\end{tabularx}
\end{table}



The details of grids utilized in the numerical methods for generating the reference solutions are presented in Table~\ref{app._tab2}. Note that in all cases, the uniform grids are employed in the temporal and spatial domain. As for the solid angular domain, the Gauss-Legendre quadrature points are employed.




\begin{table}[H]
\caption{Grid size used in the numerical method for generating reference solutions in each case. 
For the 3D  case in Sec. 3.4, we use $40\times40$ and $80\times 80$ points in the solid angular domain for $\rm{Kn}=0.1$, 1  and 10, respectively. 
}\label{app._tab2}
\newcolumntype{C}{>{\centering\arraybackslash}X}
\begin{tabularx}
{\textwidth}{m{3cm}m{2cm}m{2.5cm}m{3.5cm}}
\toprule
   & time ($t$)	& space ($\bm{x}$) & solid angular domain ($\bm{s}$)  \\
\midrule
Sec. 3.1		& - & 40& 100 \\
Sec. 3.2		& 1600 & 80 & 100 \\
Sec. 3.3		& - & $40\times 40$& $40\times 40$ \\
Sec. 3.4		& - & $40\times40\times 40$  & $40\times 40$, $80\times 80$ \\
Appendix B  & - & 40 & 100          \\
\bottomrule
\end{tabularx}
\end{table}

\section{Ablation study}
\label{sec:appendix_c}
\begin{figure}[H]
\centering
\subfigure[]{\includegraphics[width=0.3\textwidth]{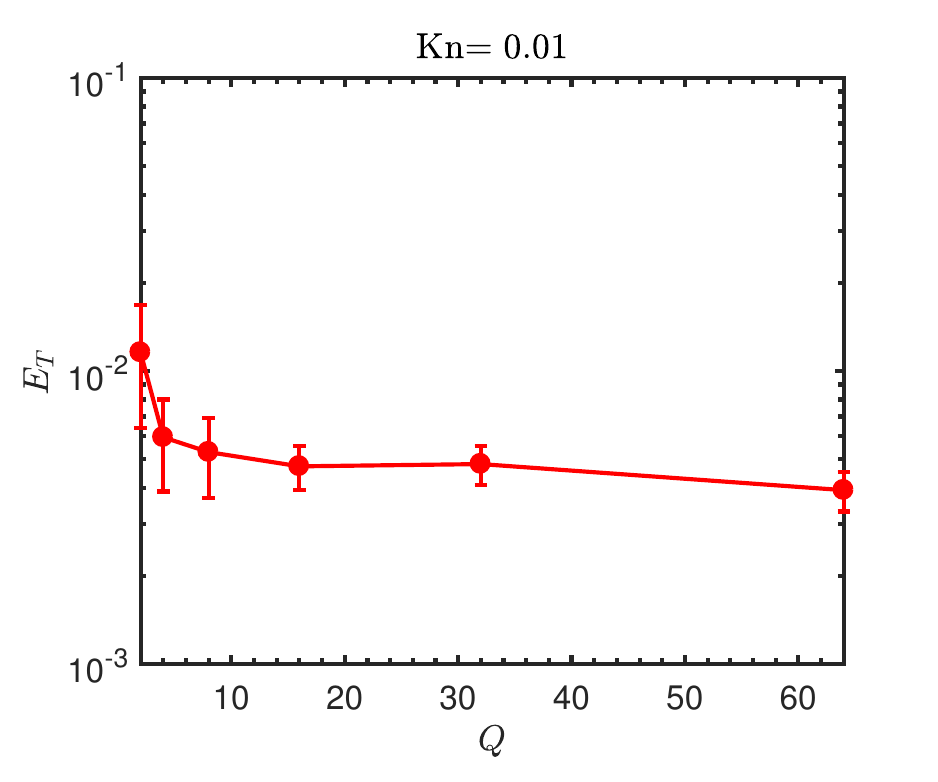}}
\subfigure[]{\includegraphics[width=0.3\textwidth]{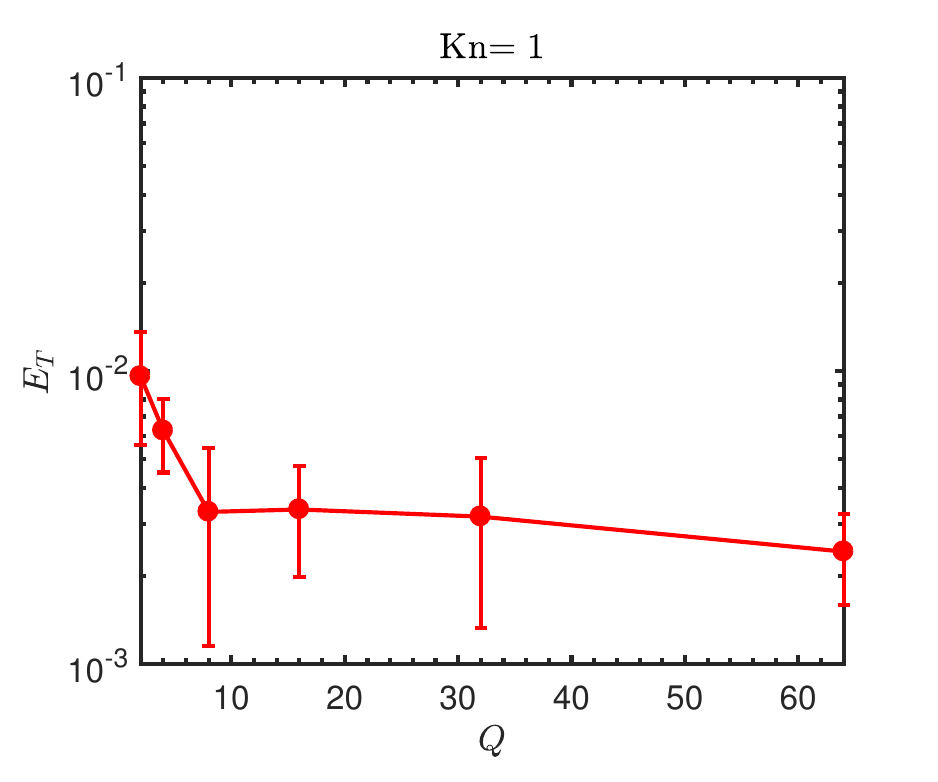}}
\subfigure[]{\includegraphics[width=0.3\textwidth]{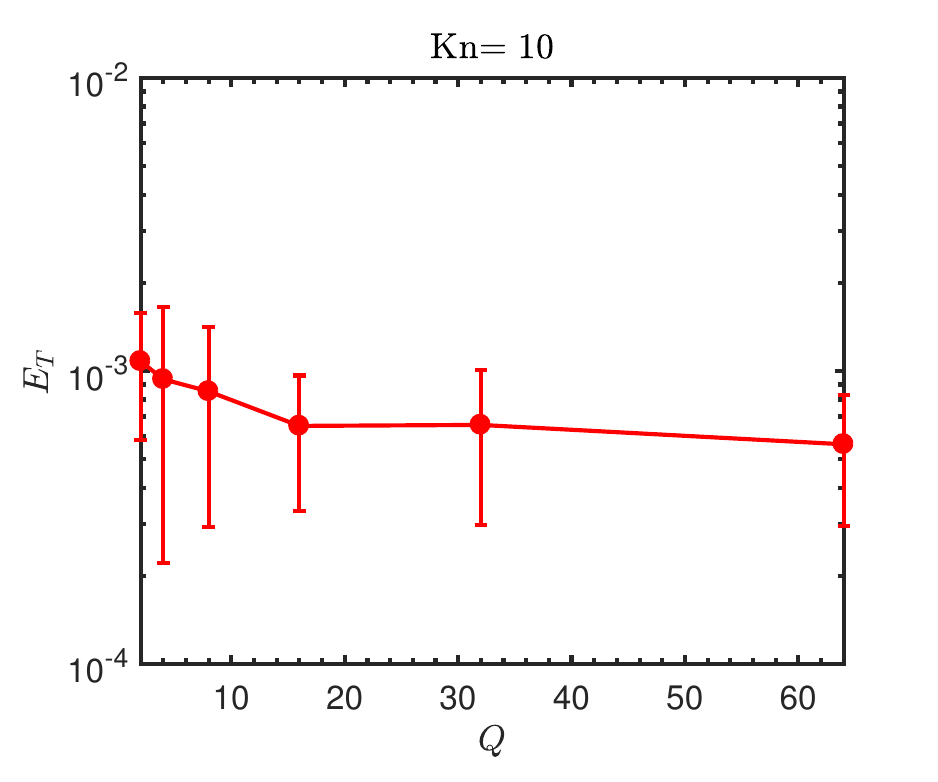}}
\caption{Quasi-1D film heat conduction: Relative $L_2$ errors of the temperature with different numbers of points in the solid angular space at each training step in MC-PINNs. (a) $\rm{Kn}=0.01$, (b) $\rm{Kn=1}$ and (c) $\rm{Kn=10}$. Error bars in each figure represent the standard deviation calculated from 5 independent runs with different random seeds. 
}\label{FIG:errorbar}
\end{figure}

To study the effect of the number of the points in the solid angular space at each training step on the predicted accuracy of MC-PINNs, we further conduct the following experiments based on two specific cases in Sec. \ref{sec:1d}, i.e., Kn $= 1$ and $10$. In each case, we test 6 different numbers of the points in the solid angular space, i.e. $B_{\bm{s}} = 2, 4, 8, 16, 32$ and 64 points at each training step. Also, we independently run MC-PINNs 5 times with different initializations for the hyperparameters to test the robustness of the MC-PINNs. The number of $B_{\bm{x}}$ is set to $40$, which is consistent with the case in Sec.~\ref{sec:1d}. The predicted temperature from MC-PINNs is depicted in Fig. \ref{FIG:errorbar}. It is observed that: (1) the computational errors in general decrease with the increase of $B_{\bm{s}}$, and (2) the relative $L_2$ errors between the results from MC-PINNs and the reference solutions in each case are below $1\%$ as $B_{\bm{s}} > 2$. In other words,  we are able to achieve good accuracy as we randomly sample more than 2 points in the solid angular domain 
at each training step in MC-PINNs for these two specific cases. 


\begin{figure}[H]
    \centering
    \includegraphics[width=0.5\textwidth]{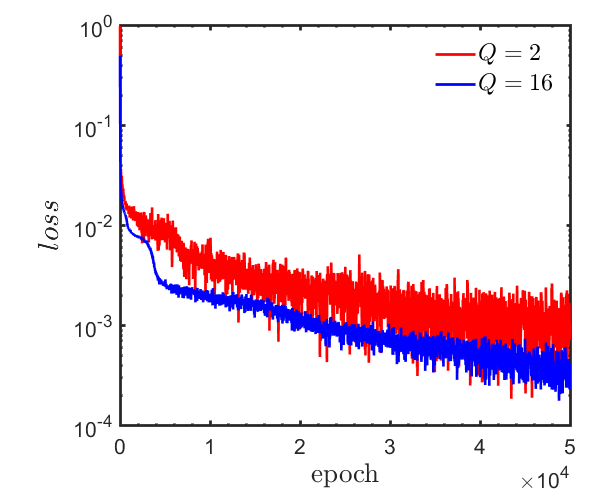}
    \caption{ Loss history for $B_{\bm{s}} = 2$ and $B_{\bm{s}}=16$ at $\rm{Kn = 1}$.}
    \label{fig:loss}
\end{figure}

We further present the loss histories for the representative case $\rm{Kn}=1.0$ with $B_{\bm{s}} = 2$ and $16$  in Fig. \ref{fig:loss}.
As shown, the loss for MC-PINNs with 16 random points in the solid angular space after 50,000 training steps is smaller than that with 2 random points in the solid angular domain. Also, the fluctuations in the loss are smaller for the case with a larger number of points in the solid angular space at each training step.


 \bibliographystyle{elsarticle-num} 
 \bibliography{refs}

\end{document}